\documentclass[prd,superscriptaddress,twocolumn,showpacs,amsmath,amssymb]{revtex4}
\usepackage{graphicx}
\usepackage{color}
\usepackage{dcolumn}
\usepackage{bm}
\usepackage{slashed}
\begin{document}
\title{Note on the gravitational and electromagnetic radiation for the Einstein-Maxwell equations with cosmological constant}

\author{Xiaokai He}
\email{hexiaokai77@163.com} \affiliation{School of Mathematics and
Computational Science, Hunan First Normal University, Changsha,
Hunan
410205,P.R.China\\
Department of Physics, Hunan Normal University, Changsha, Hunan
410081, P.R.China}

\author{Zhoujian Cao \footnote{corresponding author}}
\email{zjcao@amt.ac.cn} \affiliation{Institute of Applied
Mathematics, Academy of Mathematics and Systems Science, Chinese
Academy of Sciences, Beijing 100190, China}

\author{Jiliang Jing}
\email{jljing@hunn.edu.cn} \affiliation{Department of Physics, Key
Laboratory of Low Dimensional Quantum Structures and Quantum Control
of Ministry of Education, and Synergetic Innovation Center for
Quantum Effects and Applications, Hunan Normal University, Changsha,
Hunan 410081, P. R. China}

\begin{abstract}
In the middle of last century, Bondi and his coworkers proposed an
out going boundary condition for the Einstein equations. Based on such
boundary condition the authors theoretically solved the puzzle of
the existence problem of gravitational wave. Currently many works on
gravitational wave source modeling use Bondi's result to do
the analysis including the gravitational wave calculation in
numerical relativity. Recently more and more observations imply
that the Einstein equations should be modified with an nonzero
cosmological constant. In this work we use Bondi's original method to treat the
Einstein equations with cosmological constant for theoretical curiosity.
We find that the gravitational wave does not essentially exist if Bondi's original out going
boundary condition is imposed. We find this unphysical result is due to the
non-consistency between Bondi's original out going
boundary condition and the Einstein equations with cosmological constant.
Based on theoretical analysis,
we propose an new Bondi-type out going boundary condition for the
Einstein equations with cosmological constant. With this new
boundary condition, the gravitational wave behavior for the
Einstein equations with cosmological constant is similar to the
Einstein equations without cosmological constant. In particular,
when the cosmological constant goes to zero, the behavior of the
Einstein equations without cosmological constant is recovered.
Together with gravitational wave, electromagnetic wave is also
analyzed. The boundary conditions for electromagnetic wave does not depend
on the cosmological constant.
\end{abstract}


\maketitle


\section{Introduction}
Einstein predicted gravitational wave based on general relativity
(GR) right after his discovery of GR. Due to the complexity of the
diffeomorphism in GR, people (including Einstein himself) have
argued the existence of gravitational wave for quite a long time~\cite{Ken07}. Till 1960s, Bondi and his coworkers solved this debate
\cite{BonVanMet62,Sac62}. After that, different
instruments and plans for gravitational wave detection are
constructed and proposed \cite{Sau94,LIGO92,Dan03,GonXuBai11}. The
idea and the long-range prospect of gravitational wave astronomy
have also been constructed gradually \cite{Sch86,Sch99,CorMes09}.

In the past years, people have found many evidences coming from the
cosmological microwave background \cite{deAdeBoc00,KomSmiDun11}, the
observation of supernova Ia \cite{PerAldDel98,RieFilCha98}, the gas
fraction in X-ray luminous galaxy clusters \cite{AllRapSch08}, the
baryon acoustic oscillations \cite{EisZehHog05}, the integrated
Sachs-Wolfe effect \cite{GiaScrCri08}, strong gravitational lensing
\cite{SuyMarAug10} and more, which show our present universe is in a
state of accelerated expansion. Although many plausible models are
constructed to explain such acceleration, most observations favor
the cold dark matter model with a cosmological constant ($\Lambda$-CDM model). At least
phenomenologically the Einstein equations should be modified with a
cosmological constant $\Lambda$ at cosmological scale:
\begin{equation}
R_{ab}-\frac{1}{2}Rg_{ab}+\Lambda g_{ab}=8\pi T_{ab}.
\label{eq:ECOS}
\end{equation}
Actually, we have more evidences besides accelerated expansion of
our universe which implies there is a non-zero cosmological constant
\cite{Pad03}. The observations \cite{Pla13} show that the amplitude
of the cosmological constant $\Lambda$ is about $10^{-52}$m${}^{-2}$
in geometric units ($c=G=1$).

The theoretical description of gravitational wave can be divided
into the generation part and the propagation part. In the length scale
viewpoint, the generation part corresponds to the near zone, and the
propagation part corresponds to the far zone. In the near zone, the
Coulomb type field dominates. In the far zone, the radiation field
instead dominates. For the Einstein equations without cosmological constant,
the Coulomb field is roughly bounded in the scale of $M$, where $M$ is
the mass of the gravitational wave source. Typically the wave length of the
involved gravitational wave is several $M$. So the region farther than
several $M$ can be treated as propagation part safely. Usually people use
asymptotically flat spacetime to model this far region. In the
asymptotically flat background, the generation theory of gravitational wave is constructed
in the seminal work of Bondi, Sachs and their coworkers
\cite{BonVanMet62,Sac62}. The propagation theory is the perturbation
theory respect to the Minkowski space \cite{Mag08}. It is also possible to use cosmological background to model the far zone.
In the cosmological
background for gravitational wave propagation, some techniques using geometric-optics approximation and
perturbation theory are developed \cite{Haw68,Gla71,ChaJan76,Way78}.

For the Einstein equations with cosmological constant, part of the Coulomb field is
produced by the cosmological constant. And its action range is about $\frac{1}{\sqrt{\Lambda}}$.
So in this sense, we should take the region to about $\frac{1}{\sqrt{\Lambda}}$ as the generation part. And
the generation theory should be considered to this domain. In practice, $\frac{1}{\sqrt{\Lambda}}\sim 10^{26}$m.
Comparing to the realized gravitational wave detection events GW150914 \cite{PhysRevLett.116.061102}, GW151226 \cite{PhysRevLett.116.241103}, GW170104 \cite{PhysRevLett.118.221101}, and GW170814 \cite{PhysRevLett.119.141101}, the corresponding source distances are 410Mpc$\sim 1.3\times10^{25}$m, 440Mpc$\sim 1.4\times10^{25}$m, 880Mpc$\sim 2.7\times10^{25}$m and 540Mpc$\sim 1.7\times10^{25}$m respectively.

Many studies about the gravitational waves in cosmological
background have been taken in the past decades
\cite{Haw68,Gla71,ChaJan76,Way78}. Pioneered by Hawking
\cite{Haw68}, these investigations have considered the generation
and the propagation of gravitational waves in the
Friedman-Robertson-Walker (FRW) back ground. Their results show that
the characteristic features of gravitational waves with cosmological
effect is similar to that got in \cite{BonVanMet62,Sac62}. But these
works considered only the Einstein equations without cosmological constant.
Few works including \cite{Sma78,Chr81,GeLuoSu11,AnnNgStr11,AshBonKes15} investigated
the cosmological constant issue. Their results imply that the
geometric behavior for spacetimes with vanishing cosmological
constant $\Lambda$ is much different to the behavior for
spacetimes with non-vanishing $\Lambda$ no matter how small $\Lambda$ is.

In \cite{BonVanMet62,Sac62}, Bondi and his coworkers proposed a boundary condition, which is an essential assumption in their works. Later Penrose and his coworker used conformal completion language to reexpress the result of Bondi \cite{Pen65,NewPen66}. We would like to point that the specific conformal completion is also an assumption. In the book \cite{PenRin88}, the authors have discussed the relationship between Bondi's boundary condition (expressed as one of the peeling-off properties) and the conformal completion assumption. In \cite{AshBonKes15}, the authors tried to use the conformal technique to investigate the Einstein equations with cosmological constant. They found several subtleties. We will show in current paper that Bondi's boundary condition is not consistent to the Einstein equations with cosmological constant. Instead of Bondi's out going boundary condition, we propose an new Bondi-type boundary condition for the Einstein equations with cosmological constant. Based on the new out going boundary condition, the properties of gravitational wave for the Einstein equations with cosmological constant are similar to the properties shown in \cite{BonVanMet62,Sac62}.

Throughout this paper we use geometric units $c=G=1$. The lower case
of Latin letters denote the abstract index of tensors \cite{wald84,liang00}. We take
signature convention $(-,+,+,+)$, so $\Lambda>0$ corresponds to the
accelerated expanding universe in the $\Lambda$-cold dark matter ($\Lambda$-CDM) model. But we would like to point out that our new boundary condition is valid for both $\Lambda>0$ and $\Lambda<0$.

The rest of the paper is organized as following. In the next section we will
briefly review the analysis method taken in the seminal works \cite{BonVanMet62,Sac62}.
We will take closely the steps of this method to investigate the Einstein equations with cosmological
constant in this work. Then in Sec.~\ref{sec::III} we apply Bondi's original out
going boundary condition to the Einstein equations with cosmological
constant. And the gravitational
wave generation is investigated correspondingly. In Sec.~\ref{sec::IV} we show the non-consistency
between Bondi's original out
going boundary condition and the Einstein equation with cosmological
constant. Together with the analysis, an new Bondi-type boundary condition is proposed.
After that we apply our new boundary condition to the Einstein equations with cosmological
constant and analyze the gravitational wave properties correspondingly in Sec.~\ref{sec::V}.
At last we give a
summary and discussion in Sec.~\ref{sec::VI} to close the paper.

\section{Brief review of BMS result} \label{sec::II}
As mentioned in the above Introduction section the whole spacetime involving gravitational wave can be divided into near zone and far zone. The near zone corresponds to the generation region of the gravitational wave. In the gravitational wave generation theory of \cite{BonVanMet62,Sac62} the near zone is modeled as an isolated spacetime. The authors in \cite{BonVanMet62,Sac62} mainly considered the asymptotic region of the isolated spacetime.
Related to gravitational wave sources, this `asymptotic' region corresponds to the junction part between the near zone and the far zone of the whole spacetime.

The line element of the Bondi-Sachs form reads as
\cite{BonVanMet62,Sac62,CaoHe13}
\begin{align}
ds^2&=-(Vr^{-1}e^{2\beta}-U^2r^2e^{2\gamma})du^2-2e^{2\beta}dudr\nonumber\\
&-2Ur^2e^{2\gamma}du
d\theta+r^2(e^{2\gamma}d\theta^2+e^{-2\gamma}\sin^2\theta
d\phi^2).\label{eq:BSmetric}
\end{align}
where $r$ is the luminosity parameter. $(u,r,\theta,\phi)$ is called as Bondi-Sachs coordinate.
$V$, $U$, $\gamma$ and $\beta$ are functions of the coordinates. As explained in
\cite{CaoHe13}, the metric can almost always been written in this
form in asymptotic region, where the gravitational field is weak. For simplicity we consider axisymmetric
spacetime only in this paper. But it can be generalized to generic
spacetime straightforwardly. Based on this metric form, the vacuum
Einstein equations are divided into three groups:

(i) main equations:
\begin{eqnarray}
R_{11}=R_{12}=R_{22}=R_{33}=0,
\end{eqnarray}

(ii)  trivial equation:
\begin{eqnarray}
R_{10}=0,
\end{eqnarray}

(iii)  supplementary conditions:
\begin{eqnarray}
R_{20}=R_{00}=0,\label{eq:suppBS}
\end{eqnarray}
where $R_{ab}$ is the Ricci tensor of the four dimensional metric. And we have used index $0$ to represent $u$, and index $1$, $2$, $3$ represent $r$, $\theta$, $\phi$ respectively.
In order to solve the Einstein equations in the asymptotic region, Bondi and his coworkers \cite{BonVanMet62,Sac62} proposed an outgoing boundary condition written as
\begin{align}
\gamma=\frac{c(u,\theta)}{r}+\frac{d(u,\theta)}{r^2}+\frac{h(u,\theta)}{r^3}+\cdot\cdot\cdot\label{eq:gamma}
\end{align}
In current paper we call this boundary condition Bondi's original boundary condition. In the seminal work \cite{BonVanMet62,Sac62}, some arguments are given to support this boundary condition. And it is indeed true this boundary condition is physically reasonable. But it is an assumption logically. Of course, one more assumption is that the power series respect to $r$ converge.

Plugging this boundary condition into main equation $R_{11}=0$, one
can get
\begin{align}
\beta=H(u,\theta)-\frac{c^2}{4r^2}-\frac{2cd}{3r^3}-\frac{3ch+2d^2}{4r^4}+\cdot\cdot\cdot\label{eq:beta}
\end{align}
Put Eqs.~(\ref{eq:gamma}) and (\ref{eq:beta}) into main equation
$R_{12}=0$, one can find out
\begin{align}
U(u,\theta)=&L(u,\theta)+\frac{2e^{2H}\partial
H/\partial\theta}{r}\label{eq:U1}\\
&-\frac{e^{2H}[2c\partial H/\partial\theta+2c\cot\theta+\partial
c/\partial\theta]}{r^2}+\cdot\cdot\cdot\nonumber\\
d(u,\theta)=&0.
\end{align}
Combine Eqs.~(\ref{eq:gamma}),
(\ref{eq:beta}), (\ref{eq:U1}) and main equation $R_{22}=0$, we get
\begin{align}
V=&(L\cot\theta+\partial L/\partial\theta)r^2+
e^{2H}\big{[}2\cot\theta\partial H/\partial\theta+4(\partial H/\partial\theta)^2\nonumber\\
&2\partial^2H/\partial\theta^2+1\big{]} r
+V_0(u,\theta)+\frac{V_1(u,\theta)}{r}+\cdot\cdot\cdot
\end{align}
Then the requirement of time-like property of
$(\frac{\partial}{\partial u})^a$ as $r$ goes to infinity leads to
\begin{align}
L(u,\theta)=0.\label{eq:Leq0}
\end{align}
Preserving the metric form (\ref{eq:BSmetric}), more gauge freedom exists. Take this freedom we consider the following coordinate transformation
\begin{align}
&u=a_0(\bar{u},\bar{\theta})+\frac{a_1(\bar{u},\bar{\theta})}{\bar{r}}+\cdot\cdot\cdot\\
&r=\bar{r}+\rho_0(\bar{u},\bar{\theta})+\cdot\cdot\cdot\\
&\theta=\bar{\theta}+\frac{b_1(\bar{u},\bar{\theta})}{\bar{r}}+\cdot\cdot\cdot\\
&\phi=\bar{\phi}.
\end{align}
In order to preserve the metric form (\ref{eq:BSmetric}), we have equations
\begin{eqnarray}
\bar{g}_{11}=0&\Rightarrow& 2e^{2H} a_1+b_1^2=0,\label{eq:a1}\\
\bar{g}_{12}=0&\Rightarrow& e^{2H}\frac{\partial
a_0}{\partial\bar{\theta}}+b_1=0,\label{eq:b1}\\
\bar{g}_{22}\bar{g}_{33}=\bar{r}^4\sin^2\bar{\theta}&\Rightarrow&
2\rho_0+\frac{\partial
b_1}{\partial\bar{\theta}}+b_1\cot\bar{\theta}=\frac{\partial
a_0}{\partial\bar{\theta}}\frac{\partial H}{\partial\theta}e^{2H}.\nonumber\\
\label{eq:rho0}
\end{eqnarray}
The coordinate transformation gives us
\begin{eqnarray}
\bar{g}_{01}=\frac{\partial a_0}{\partial
\bar{u}}e^{2H}+O(\bar{r}^{-1}).
\end{eqnarray}
If we require
\begin{eqnarray}
\frac{\partial a_0}{\partial \bar{u}}=e^{-2H},\label{eq:odea0}
\end{eqnarray}
we can get
\begin{eqnarray}
\bar{H}=0.
\end{eqnarray}
Firstly we choose a smooth function with argument $\bar{\theta}$. Use this function as an initial data to solve the ordinary differential equation (\ref{eq:odea0}) to get $a_0(\bar{u},\bar{\theta})$. Plug this $a_0$ into Eq.~(\ref{eq:b1}) to get $b_1(\bar{u},\bar{\theta})$. Insert this $b_1$ into Eq.~(\ref{eq:a1}) to get $a_1(\bar{u},\bar{\theta})$. And combine the solutions $a_0$ and $b_1$ with Eq.~(\ref{eq:rho0}) we can get $\rho_0(\bar{u},\bar{\theta})$. With these functions, we have a concrete coordinate transformation. With this new coordinate system, we still have the Bondi-Sachs metric form (\ref{eq:BSmetric}). And at the same time the equations (\ref{eq:gamma})-(\ref{eq:Leq0}) keep the same form but with vanishing $\bar{H}$. In the following discussion we will use this coordinate system. But for notation simplicity, we drop the over bar.

In summary we can write down the solution of the Einstein equations as
\begin{align}
&\gamma=\frac{c(u,\theta)}{r}+\frac{h(u,\theta)}{r^3}+\cdot\cdot\cdot\label{eq:solgamma}\\
&\beta=-\frac{c^2}{4r^2}-\frac{3ch}{4r^4}+\cdot\cdot\cdot\label{eq:solbeta}\\
&U=-\frac{2c\cot\theta+\partial
c/\partial\theta}{r^2}\nonumber\\
&+\frac{2N(u,\theta)+3c
\frac{\partial
c}{\partial\theta}+4c^2\cot\theta}{r^3}+\cdot\cdot\cdot\label{eq:solU}\\
&V=r-2M(u,\theta)+\frac{V_1(u,\theta)}{r}+\cdot\cdot\cdot\label{eq:solV}
\end{align}
Submitting these solutions into supplementary conditions (\ref{eq:suppBS}), we get
\begin{eqnarray}
&\frac{\partial M}{\partial u}=-\frac{\partial c}{\partial
u}+\frac{1}{2}\frac{\partial ^3c}{\partial\theta^2\partial
u}+\frac{3}{2}\frac{\partial c}{\partial\theta\partial
u}\cot\theta-(\frac{\partial c}{\partial u})^2,\\
&\frac{\partial N}{\partial u}=-\frac{1}{3}\bigg{[}\frac{\partial
M}{\partial \theta}+3c\frac{\partial ^2c}{\partial \theta\partial
u}+4c\frac{\partial c}{\partial u}\cot\theta+\frac{\partial
c}{\partial u}\frac{\partial c}{\partial \theta}\bigg{]}.
\end{eqnarray}

We would like to use Newman-Penrose scalar
$\Psi_4$ to represent the gravitational wave like what is done in numerical relativity.
For this purpose, we choose the following quasi-Kinnersley tetrad:
\begin{align}
&l^a=(\frac{\partial}{\partial r})^a,\label{eq:QKtetradl}\\
&n^a=e^{-2\beta}(\frac{\partial}{\partial
u})^a-\frac{Ve^{-2\beta}}{2r}(\frac{\partial}{\partial
r})^a+Ue^{-2\beta}(\frac{\partial}{\partial\theta})^a,\label{eq:QKtetradn}\\
&m^a=\frac{1}{\sqrt{2}r}[e^{-\gamma}(\frac{\partial}{\partial\theta})^a+\frac{i
e^{\gamma}}{\sin\theta}(\frac{\partial}{\partial\phi})^a].\label{eq:QKtetradm}
\end{align}
The resulted $\Psi_4$ reads as
\begin{align}
\Psi_4=-\frac{\partial^2c/\partial u^2}{r}+\cdot\cdot\cdot
\end{align}
$\Psi_4$ is related to geodesic deviation equation which is the
basis of gravitational wave detection.

Plug the solutions (\ref{eq:solgamma})-(\ref{eq:solV}) in the Einstein equations, we can find an interesting structure of the solution. $c(u,\theta)$ is completely free. Other coefficients like $h$, $V_0$, $V_1$ are controlled by a corresponding ordinary differential equation, which depends on $c(u,\theta)$. These ordinary differential equations are respect to $u$, which is called Bondi time. So the other coefficients than $c$ have only freedom of some initial data. These initial data are some smooth functions of $\theta$. And they can be freely chosen. This interesting structure corresponds to the characteristic initial value problem \cite{CaoHe13}.

So we can see that the gravitational wave is fully encoded in function $c$. And this $c$ determines all other coefficients. In this sense, $c$ is called Bondi's news function.

\section{Apply Bondi's original boundary condition to the Einstein equations with $\Lambda$} \label{sec::III}
In the last section, we have reviewed the analysis method taken by Bondi and his coworkers.
In this section we will follow this method closely to investigate the Einstein equations
with non-vanishing cosmological constant $\Lambda$. And we apply the
boundary condition (\ref{eq:gamma}) to this problem.
In order to compare the behavior of gravitational wave and electromagnetic wave, we treat the Einstein equations and the Maxwell equations together. We consider the coupled Maxwell equations and the Einstein equations
\begin{eqnarray}
R_{ab}-\frac{1}{2}Rg_{ab}+\Lambda g_{ab}&=&8\pi T_{ab},\label{EE1}\\
\nabla^aF_{ab}&=&0,\\
\nabla_{[a}F_{bc]}&=&0,
\end{eqnarray}
where $T_{ab}$ denotes the energy-momentum tensor of
electromagnetic field and it is given by
\begin{eqnarray}
T_{ab}=\frac{1}{4\pi}(F_{ac}F_{b}^{\
c}-\frac{1}{4}g_{ab}F_{cd}F^{cd})
\end{eqnarray}
The Eq.~(\ref{EE1}) can be reexpressed as
\begin{eqnarray}\label{EE2}
R_{ab}=8\pi(T_{ab}-\frac{1}{2}g_{ab}T)+\Lambda
g_{ab}
\end{eqnarray}
where $T$ is the trace of energy-momentum tensor. For pure electromagnetic field, we have
\begin{eqnarray}
T=g^{ab}T_{ab}=\frac{1}{4\pi}(F_{ab}F^{ab}-F_{ab}F^{ab})=0.
\end{eqnarray}
Then the Einstein filed equations are simplified as
\begin{eqnarray}\label{EE3}
R_{ab}=8\pi T_{ab}+\Lambda g_{ab}.
\end{eqnarray}

As last section, we start from the Bondi-Sachs metric form (\ref{eq:BSmetric}). And the axisymmetric assumption is also taken for simplicity. Then the electromagnetic field
takes the following form in Bondi-Sachs coordinates
\begin{eqnarray}
F=F_{01}du\wedge dr+F_{02}du\wedge d\theta+F_{12}dr\wedge d\theta
\end{eqnarray}
And the Maxwell equations become
\begin{eqnarray}
&&r^3\frac{\partial F_{01}}{\partial u}+r^3F_{01}\frac{\partial
U}{\partial\theta}-2r^3UF_{01}\frac{\partial\beta}{\partial\theta}-2r^3F_{01}\frac{\partial\beta}{\partial
u}\nonumber\\
&&-r^5Ue^{2\gamma-2\beta}F_{01}\frac{\partial U}{\partial
r}+r^5U^2e^{2\gamma-2\beta}F_{12}\frac{\partial
U}{\partial r}\nonumber\\
&&+r^3U\frac{\partial F_{01}}{\partial \theta}-r^2V\frac{\partial
F_{01}}{\partial
r}-2rVF_{01}+r^3UF_{01}\cot\theta\nonumber\\
&&-re^{2\beta-2\gamma}\frac{\partial
F_{02}}{\partial\theta}-re^{2\beta-2\gamma}F_{02}\cot\theta+3rUVF_{12} \nonumber\\
&&+2re^{2\beta-2\gamma}F_{02}\frac{\partial\gamma}{\partial\theta}+r^3U\frac{\partial
F_{02}}{\partial r}+e^{2\beta-2\gamma}F_{12}\frac{\partial
V}{\partial\theta}\nonumber\\
&& +2r^3UF_{12}\frac{\partial\beta}{\partial
u}-r^2UF_{12}\frac{\partial V}{\partial r}-2r^3UF_{12}\frac{\partial
\gamma}{\partial
u}\nonumber\\
&&-2r^3UF_{12}\frac{\partial
U}{\partial\theta}-2r^3U^2F_{12}\frac{\partial\gamma}{\partial
\theta}+2r^3U^2F_{12}\frac{\partial\beta}{\partial\theta}\nonumber\\
&&-2r^2UVF_{12}\frac{\partial\beta}{\partial
r}+r^2VF_{12}\frac{\partial U}{\partial
r}+2r^2UVF_{12}\frac{\partial\gamma}{\partial
r}\nonumber\\
&&-r^3F_{12}\frac{\partial U}{\partial
u}-2r^3UF_{02}\frac{\partial\gamma}{\partial
r}+2r^2VF_{01}\frac{\partial\beta}{\partial r}=0,\label{maxwell1}\\
&&-r^2\frac{\partial F_{01}}{\partial
r}-2rF_{01}+2r^2F_{01}\frac{\partial \beta}{\partial
r}+r^2F_{12}\frac{\partial U}{\partial
r}\nonumber\\
&&+2rF_{12}U+r^2U\frac{\partial F_{12}}{\partial
r}-2r^2UF_{12}\frac{\partial\beta}{\partial
r}-e^{2\beta-2\gamma}\frac{\partial F_{12}}{\partial
\theta}\nonumber\\
&&+2e^{2\beta-2\gamma}F_{12}\frac{\partial\gamma}{\partial
\theta}-F_{12}e^{2\beta-2\gamma}\cot\theta=0,\label{maxwell2}\\
&&-2r^2F_{02}\frac{\partial\gamma}{\partial r}+r^2\frac{\partial
F_{02}}{\partial r}-r^4e^{2\gamma-2\beta}F_{01}\frac{\partial
U}{\partial
r}+F_{12}V\nonumber\\
&&-2r^2F_{12}\frac{\partial \gamma}{\partial
r}+r^2UF_{12}\cot\theta+r^2\frac{\partial F_{12}}{\partial
u}\nonumber\\
&&+2rVF_{12}\frac{\partial\gamma}{\partial
r}-2r^2UF_{12}\frac{\partial\gamma}{\partial \theta}
+r^4Ue^{2\gamma-2\beta}F_{12}\frac{\partial U}{\partial r}\nonumber\\
&&-rF_{12}\frac{\partial V}{\partial r} -rV\frac{\partial
F_{12}}{\partial r}+r^2U\frac{\partial F_{12}}{\partial
\theta}=0,\label{maxwell3}\\
&&\frac{\partial F_{12}}{\partial u}-\frac{\partial F_{02}}{\partial
r}+\frac{\partial F_{01}}{\partial \theta}=0.\label{maxwell4}
\end{eqnarray}

Regarding to the Einstein equations, they are divided into three groups as last section:

(i) main equations:
\begin{eqnarray}
&&R_{11}=8\pi T_{11}+\Lambda g_{11},\hspace{2mm}R_{12}=8\pi
T_{12}+\Lambda g_{12},\hspace{2mm}\\
&&R_{22}=8\pi T_{22}+\Lambda g_{22},\hspace{2mm}R_{33}=8\pi
T_{33}+\Lambda g_{33}.
\end{eqnarray}

(ii) trivial equation:
\begin{eqnarray}
R_{10}=8\pi T_{10}+\Lambda g_{10}.
\end{eqnarray}

(iii) supplementary conditions:
\begin{eqnarray}
R_{20}=8\pi T_{20}+\Lambda g_{20},\hspace{2mm}R_{00}=8\pi
T_{00}+\Lambda g_{00}.\label{supplambda}
\end{eqnarray}

The explicit
expressions of the main equations take the following form.
\begin{eqnarray}
&&\hspace{15mm}R_{11}=8\pi T_{11}+\Lambda g_{11}
\hspace{3mm}\Rightarrow\hspace{3mm}
\nonumber\\
&&(\frac{\partial\gamma}{\partial
r})^2-2\frac{\partial\beta}{\partial
r}+\frac{F_{12}^2}{re^{2\gamma}}=0.\label{main1}\\
&&\hspace{15mm}R_{12}=8\pi T_{12}+\Lambda g_{12} \hspace{3mm}\Rightarrow\nonumber\\
&&-4e^{2\beta}\frac{\partial\beta}{\partial\theta}-4r\cot\theta
e^{2\beta}\frac{\partial\gamma}{\partial
r}-2re^{2\beta}\frac{\partial^2\gamma}{\partial
r\partial\theta}\nonumber\\
&&-2r^3e^{2\gamma}\frac{\partial\gamma}{\partial r}\frac{\partial
U}{\partial r}-4r^2e^{2\gamma}\frac{\partial U}{\partial
r}+2r^3e^{2\gamma}\frac{\partial\beta}{\partial r}\frac{\partial
U}{\partial
r}\nonumber\\
&&+2re^{2\beta}\frac{\partial^2\beta}{\partial\theta\partial
r}-r^3e^{2\gamma}\frac{\partial^2U}{\partial
r^2}+4re^{2\beta}\frac{\partial\gamma}{\partial\theta}\frac{\partial\gamma}{\partial
r}\nonumber\\
&&+4rF_{12}(F_{12}U-F_{01})=0.\label{main2}\\
&&\hspace{16mm}R_{22}=8\pi T_{22}+\Lambda g_{22}\hspace{2mm}\Rightarrow\nonumber\\
&&-2rU\cot\theta-6e^{2\beta-2\gamma}\frac{\partial\gamma}{\partial
\theta}\cot\theta-2r^2U\frac{\partial\gamma}{\partial
r}\cot\theta\nonumber\\
&&+4e^{2\beta-2\gamma}(\frac{\partial\beta}{\partial\theta})^2+4e^{2\beta-2\gamma}\frac{\partial^2\beta}{\partial\theta^2}+4e^{2\beta-2\gamma}(\frac{\partial\gamma}{\partial\theta})^2\nonumber\\
&&-2e
^{2\beta-2\gamma}\frac{\partial^2\gamma}{\partial\theta^2}+2r\frac{\partial
V}{\partial r}\frac{\partial\gamma}{\partial r}-2r^2\frac{\partial
U}{\partial\theta}\frac{\partial\gamma}{\partial
r}-2r^2\frac{\partial U}{\partial r}\frac{\partial\gamma}{\partial
\theta}\nonumber\\
&&-4rU\frac{\partial\gamma}{\partial\theta}-4r^2U\frac{\partial^2\gamma}{\partial\theta\partial
r}+2rV\frac{\partial^2\gamma}{\partial r^2}+2\frac{\partial
V}{\partial
r}\nonumber\\
&&-4e^{2\beta-2\gamma}\frac{\partial\beta}{\partial\theta}\frac{\partial\gamma}{\partial\theta}
+r^4e^{2\gamma-2\beta}(\frac{\partial U}{\partial
r})^2-6r\frac{\partial
U}{\partial\theta}+2V\frac{\partial\gamma}{\partial
r}\nonumber\\
&&-4r^2\frac{\partial ^2\gamma}{\partial u\partial
r}-2r^2\frac{\partial^2U}{\partial\theta\partial
r}-2e^{2\beta-2\gamma}-4e^{-2\gamma}F_{02}F_{12}\nonumber\\
&&+2\frac{V}{r}e^{-2\gamma}F_{12}^2+2r^2e^{-2\beta}F_{01}^2
-4r^2e^{-2\beta}UF_{01}F_{12}\nonumber\\&&-4r\frac{\partial
\gamma}{\partial u}+2r^2U^2e^{-2\beta}F_{12}^2+2\Lambda
r^2e^{2\beta}=0.\label{main3}
\end{eqnarray}

\begin{eqnarray}
&&\hspace{3cm}R_{33}=8\pi T_{33}+\Lambda g_{33}\hspace{2mm}\Rightarrow\nonumber\\
&&-re^{2\gamma}\frac{\partial
U}{\partial\theta}+r^2e^{2\gamma}\frac{\partial U}{\partial
r}\frac{\partial\gamma}{\partial\theta}+2rUe^{2\gamma}\frac{\partial\gamma}{\partial\theta}+2re^{2\gamma}\frac{\partial\gamma}{\partial
u}\nonumber\\
&&-e^{2\gamma}V\frac{\partial\gamma}{\partial
r}+e^{2\gamma}\frac{\partial V}{\partial
r}-e^{2\beta}-3e^{2\beta}\frac{\partial\gamma}{\partial\theta}\cot\theta
\nonumber\\
&&+2e^{2\beta}\frac{\partial\beta}{\partial\theta}\cot\theta
-2e^{2\beta}\frac{\partial\gamma}{\partial\theta}\frac{\partial\beta}{\partial\theta}+r^2Ue^{2\gamma}\frac{\partial\gamma}{\partial
r}\cot\theta\nonumber\\
&&-3rUe^{2\gamma}\cot\theta+r^2e^{2\gamma}\frac{\partial\gamma}{\partial
r}\frac{\partial U}{\partial \theta}-r^2e^{2\gamma}\frac{\partial
U}{\partial r}\cot\theta\nonumber\\
&&-re^{2\gamma}\frac{\partial V}{\partial r}\frac{\partial
\gamma}{\partial
r}+2Ur^2e^{2\gamma}\frac{\partial^2\gamma}{\partial\theta\partial
r}+2r^2e^{2\gamma}\frac{\partial^2\gamma}{\partial u\partial
r}\nonumber\\
&&-rVe^{2\gamma}\frac{\partial^2\gamma}{\partial
r^2}+2e^{2\beta}(\frac{\partial\gamma}{\partial\theta})^2-e^{2\beta}\frac{\partial^2\gamma}{\partial\theta^2}\nonumber\\
&&
+r^2e^{2\gamma-2\beta}F_{01}^2-2r^2Ue^{2\gamma-2\beta}F_{01}F_{12}+{2}F_{12}F_{02}\nonumber\\
&&-\frac{1}{r}VF_{12}^2+r^2U^2e^{2\gamma-2\beta}F_{12}^2+\Lambda
r^2e^{2\gamma+2\beta}=0\label{main4}
\end{eqnarray}
Aided with Eq.~(\ref{main4}), the Eq.~(\ref{main3}) reduces to
\begin{eqnarray}
&&2e^{2\gamma-2\beta}\frac{\partial V}{\partial
r}-r^2e^{2\gamma-2\beta}\frac{\partial U}{\partial r}\cot\theta-2
-6\frac{\partial \gamma}{\partial\theta}\cot\theta\nonumber\\
&&+2(\frac{\partial
\beta}{\partial\theta})^2+4(\frac{\partial\gamma}{\partial\theta})^2+2\frac{\partial^2\beta}{\partial\theta^2}-2\frac{\partial^2\gamma}{\partial\theta^2}
+2\frac{\partial\beta}{\partial\theta}\cot\theta\nonumber\\
&&+\frac{1}{2}r^4e^{4\gamma-4\beta}(\frac{\partial U}{\partial
r})^2-4re^{2\gamma-2\beta}U\cot\theta-4\frac{\partial\beta}{\partial\theta}\frac{\partial\gamma}{\partial\theta}
\nonumber\\
&&-4re^{2\gamma-2\beta}\frac{\partial
U}{\partial\theta}-r^2e^{2\gamma-2\beta}\frac{\partial ^2U}{\partial
r\partial\theta}+2r^2e^{2\gamma-4\beta}(F_{01}^2\nonumber\\
&&-2F_{01}F_{12}U+U^2F_{12}^2)+2\Lambda r^2
e^{2\gamma}=0\label{main5}
\end{eqnarray}

For both $\gamma$ and $F_{12}$ we take the assumption of Bondi's original out going boundary condition
\begin{eqnarray}
\gamma=\frac{c(u,\theta)}{r}+\frac{d(u,\theta)}{r^2}+\frac{h(u,\theta)}{r^3}+\cdot\cdot\cdot\label{eq:gamma2}\\
F_{12}=\frac{A_1(u,\theta)}{r}+\frac{A_2(u,\theta)}{r^2}+\frac{A_3(u,\theta)}{r^3}+\cdot\cdot\cdot\label{eq:f12}
\end{eqnarray}

Plugging this boundary condition into Eq.~(\ref{maxwell2}) we get
\begin{eqnarray}
\frac{\partial A_1(u,\theta)}{\partial\theta}+A_1\cot\theta=0.\label{com_1}
\end{eqnarray}
Integrating it respect to $\theta$, we get
\begin{eqnarray}
A_1(u,\theta)=\frac{\Sigma(u)}{\sin\theta}
\end{eqnarray}
where $\Sigma(u)$ is the integration constant. The
regularity condition for $A_1(u,\theta)$ on the sphere of constant $u$ requires $\Sigma(u)=0$, which implies
$A_1(u,\theta)=0$.

Combine this result and the above boundary conditions with
the main equation $(\ref{main1})$, we get
\begin{eqnarray}
\beta=H(u,\theta)-\frac{c^2}{4r^2}-\frac{2cd}{3r^3}-\frac{(A_2)^2+6ch+4d^2}{8r^4}+\cdot\cdot\cdot
\end{eqnarray}

Then we analyze the Maxwell equation (\ref{maxwell2}) again and obtain
\begin{eqnarray}
&&F_{01}=\frac{B_2(u,\theta)}{r^2}+\frac{1}{r^3}\bigg{[}2\cdot(A_2)e^{2H}\frac{\partial
H}{\partial\theta}\nonumber\\
&&+LA_3+e^{2H}\cdot\frac{\partial
A_2}{\partial\theta}+e^{2H}(A_2)\cot\theta\bigg{]}+\frac{B_4(u,\theta)}{r^4}+\cdot\cdot\cdot\nonumber\\
\end{eqnarray}
At the same time the main equation $(\ref{main2})$ gives us
\begin{align}
U(u,\theta)=&L(u,\theta)+\frac{2e^{2H}\partial
H/\partial\theta}{r}\nonumber\\
&-\frac{e^{2H}[2c\partial H/\partial\theta+2c\cot\theta+\partial
c/\partial\theta]}{r^2}+\cdot\cdot\cdot\label{eq:U}
\end{align}
and
\begin{eqnarray}
\frac{\partial d(u,\theta)}{\partial\theta}+2
d(u,\theta)\cot\theta=0.\label{eq:58}
\end{eqnarray}
Integrating (\ref{eq:58}) we get
\begin{eqnarray}
d(u,\theta)=\frac{\Delta(u)}{\sin^2\theta},
\end{eqnarray}
where $\Delta(u)$ is the integration constant. Similar to above $A_1$, the
regularity conditions lead $\Delta(u)=0$, which implies
$d(u,\theta)=0$.

Combine these results with the main equation $(\ref{main5})$ we get
\begin{eqnarray}
&&V=-\frac{1}{3}e^{2H}\Lambda
r^3+(L\cot\theta+L_2)r^2+e^{2H}\big{[}1+4(\frac{\partial
H}{\partial\theta})^2\nonumber\\
&& +2\frac{\partial ^2H}{\partial\theta^2}+2\frac{\partial
H}{\partial\theta}\cot\theta+\frac{1}{2}\Lambda c^2\big{]}r
+V_0(u,\theta)+\cdot\cdot\cdot
\end{eqnarray}

Different to the case in the last section, the existence of the $r^3$
term in $V$ prevent us to deduce $L$ vanishes based on the requirement of time-like property of $(\frac{\partial}{\partial u})^a$.
But we can still find a coordinate transformation among the BS coordinates to
let $H$ vanish with following procedure. Start from some coordinate with non zero $H$, we
introduce a coordinate transformation as following
\begin{align}
&u=a_0(\bar{u},\bar{\theta})+\frac{a_1(\bar{u},\bar{\theta})}{\bar{r}}+\cdot\cdot\cdot\label{ctu}\\
&r=K(\bar{u},\bar{\theta})\bar{r}+\rho_0(\bar{u},\bar{\theta})+\cdot\cdot\cdot\label{ctr}\\
&\theta=b_0(\bar{u},\bar{\theta})+\frac{b_1(\bar{u},\bar{\theta})}{\bar{r}}+\cdot\cdot\cdot\label{cttheta}\\
&\phi=\bar{\phi}.\label{ctphi}
\end{align}
After this coordinate transformation, we have an new expansion of
$\beta$ with
\begin{align}
&\bar{H}=[(e^{4H}\frac{\Lambda}{3}+L^2)Ka_1+e^{2H}-LKb_1]K\frac{\partial
a_0}{\partial\bar{u}}\nonumber\\
&\hspace{10mm} +(b_1-La_1)K^2\frac{\partial b_0}{\partial\bar{u}}-1.\label{H5old}
\end{align}
In order to let the bared coordinate belongs to BS coordinate also, we need more equations
\begin{align}
&\bar{g}_{11}=0,\label{eq:11}\\
&\bar{g}_{12}=0,\label{eq:12}\\
&\bar{g}_{22}\bar{g}_{33}-\bar{r}^4\sin^2\bar{\theta}=0,\label{eq:2233}\\
&\bar{g}_{22}=\bar{r}^2+\mathcal{O}(\bar{r}),\label{eq:22}\\
&\bar{g}_{33}=\bar{r}^2\sin^2\bar{\theta}+\mathcal{O}(\bar{r}).\label{eq:33}
\end{align}
Note that the equations (\ref{eq:22}) and (\ref{eq:33}) guarantee
the leading order of the condition (\ref{eq:2233}). While the higher order terms of Eqs.~(\ref{eq:11})-(\ref{eq:2233}) affect the corresponding higher order coefficients of the coordinate transformation, the coefficients $a_0$, $a_1$, $K$, $b_0$ and $b_1$ are independent of these higher order terms. For the coefficients we concern about we have
\begin{align}
&[e^{4H}\frac{\Lambda}{3}+L^2]K^2a_1^2+2e^{2H}Ka_1-2LK^2a_1b_1+K^2b_1^2=0,\label{eq:group1a}\\
&[e^{4H}\frac{\Lambda}{3}Ka_1+L^2Ka_1-e^{2H}+LKb_1]K\frac{\partial
a_0}{\partial\bar{\theta}}\nonumber\\
&+(La_1-b_1)K^2\frac{\partial b_0}{\partial\bar{\theta}}=0,\label{eq:group1c}\\
&[e^{4H}\frac{\Lambda}{3}+L^2]K^2(\frac{\partial
a_0}{\partial\bar{\theta}})^2-2LK^2\frac{\partial
a_0}{\partial\bar{\theta}}\frac{\partial
b_0}{\partial\bar{\theta}}+K^2(\frac{\partial
b_0}{\partial\bar{\theta}})^2=1,\\
&K^2\sin^2(b_0)=\sin^2(\bar{\theta}).\label{eq:group1b}
\end{align}
Solving the algebra equations
(\ref{eq:group1a}), (\ref{eq:group1c}) and (\ref{eq:group1b}), we can get
$K$, $a_1$ and $b_1$ in terms of $a_0$, $b_0$, $\frac{\partial
a_0}{\partial\bar{\theta}}$, $\frac{\partial
b_0}{\partial\bar{\theta}}$. Submitting these solutions into (\ref{eq:group1c}) and $\tilde{H}=0$ we get equations for $a_0$ and $b_0$ with the following form
\begin{align}
&F_1(\bar{u},\bar{\theta},a_0,b_0,\frac{\partial a_0}{\partial \bar{\theta}},\frac{\partial
b_0}{\partial \bar{\theta}})\frac{\partial a_0}{\partial
\bar{u}}\nonumber\\
&+F_2(\bar{u},\bar{\theta},a_0,b_0,\frac{\partial a_0}{\partial \bar{\theta}},\frac{\partial
b_0}{\partial \bar{\theta}})\frac{\partial b_0}{\partial \bar{u}}
=1,\label{eq:F12}\\
&F_3(\bar{u},\bar{\theta},a_0,b_0)(\frac{\partial a_0}{\partial
\bar{\theta}})^2+F_4(\bar{u},\bar{\theta},a_0,b_0)\frac{\partial a_0}{\partial \bar{\theta}}\frac{\partial
b_0}{\partial \bar{\theta}}\nonumber\\
&+F_5(\bar{u},\bar{\theta},a_0,b_0)(\frac{\partial b_0}{\partial
\bar{\theta}})^2=1.\label{eq:F345}
\end{align}
Equivalently, we can consider
\begin{align}
&\frac{\partial a_0}{\partial \bar{u}}+f_1(\bar{u},\bar{\theta},a_0,b_0,\frac{\partial
a_0}{\partial \bar{\theta}},\frac{\partial b_0}{\partial \bar{\theta}})\frac{\partial
b_0}{\partial \bar{u}}\nonumber\\
&=f_2(\bar{u},\bar{\theta},a_0,b_0,\frac{\partial a_0}{\partial
\bar{\theta}},\frac{\partial
b_0}{\partial \bar{\theta}}),\label{o1}\\
&f_3(\bar{u},\bar{\theta},a_0,b_0,\frac{\partial a_0}{\partial \bar{\theta}},\frac{\partial
b_0}{\partial \bar{\theta}})=0.\label{o2}
\end{align}
To treat above resulted equations, we introduce auxiliary variables as
\begin{align}
u_1\equiv\frac{\partial a_0}{\partial \bar{u}},\ \
u_2\equiv\frac{\partial a_0}{\partial \bar{\theta}},\ \
u_3\equiv\frac{\partial b_0}{\partial \bar{\theta}}.\label{eq:def}
\end{align}
With these definitions, the arguments of $f_1$, $f_2$ and $f_3$ become $\bar{u}$, $\bar{\theta}$, $a_0$, $b_0$, $u_2$ and $u_3$. Together with Eq.~(\ref{o1}) and the time ($\bar{u}$) derivatives of Eq.~(\ref{o2}), we get
\begin{align}
&\frac{\partial b_0}{\partial
\bar{u}}=\frac{f_2-u_1}{f_1},\label{s1}\\
&\label{s2}
\frac{\partial a_0}{\partial \bar{u}}=u_1,\\
&\label{s3}
\frac{\partial u_2}{\partial \bar{u}}=\frac{\partial u_1}{\partial \bar{\theta}},\\
&\label{s4}
\frac{\partial u_3}{\partial
\bar{u}}=\frac{1}{f_1^2}\bigg{[}-(f_{1,\bar{\theta}}+f_{1,a_0}\frac{\partial
a_0}{\partial \bar{\theta}}+f_{1,b_0}\frac{\partial b_0}{\partial
\bar{\theta}}+f_{1,u_2}\frac{\partial u_2}{\partial \bar{\theta}}\nonumber\\
&+f_{1,u_3}\frac{\partial
u_3}{\partial \bar{\theta}})(f_2-u_1)+f_1\cdot(f_{2,\bar{\theta}}+f_{2,a_0}\frac{\partial
a_0}{\partial \bar{\theta}}\nonumber\\
&+f_{2,b_0}\frac{\partial b_0}{\partial
\bar{\theta}}+f_{2,u_2}\frac{\partial u_2}{\partial \bar{\theta}}+f_{2,u_3}\frac{\partial
u_3}{\partial \bar{\theta}}-\frac{\partial u_1}{\partial \bar{\theta}})\bigg{]},\\
&\label{s5}
f_{3,\bar{u}}+f_{3,a_0}u_1+f_{3,b_0}\frac{f_2-u_1}{f_1}+f_{3,u_2}\frac{\partial
u_2}{\partial \bar{u}}+f_{3,u_3}\frac{\partial u_3}{\partial \bar{u}}=0,
\end{align}
where the notations like $f_{1,\bar{u}}$ mean the partial derivative of the specified $f$ respect to the specified argument. Replace $\frac{\partial u_3}{\partial \bar{u}}$ in (\ref{s5}) with the right hand side of (\ref{s4}) we get
\begin{align}
&f_{3,\bar{u}}+f_{3,a_0}u_1+f_{3,b_0}\frac{f_2-u_1}{f_1}+f_{3,u_2}\frac{\partial
u_2}{\partial t}\nonumber\\
&+\frac{f_{3,u_3}}{f_1^2}\bigg{[}-(f_{1,\bar{\theta}}+f_{1,a_0}\frac{\partial
a_0}{\partial \bar{\theta}}+f_{1,b_0}\frac{\partial b_0}{\partial
\bar{\theta}}+f_{1,u_2}\frac{\partial u_2}{\partial \bar{\theta}}\nonumber\\
&+f_{1,u_3}\frac{\partial
u_3}{\partial \bar{\theta}})(f_2-u_1)+f_1\cdot(f_{2,\bar{\theta}}+f_{2,a_0}\frac{\partial
a_0}{\partial \bar{\theta}}\nonumber\\
&+f_{2,b_0}\frac{\partial b_0}{\partial
\bar{\theta}}+f_{2,u_2}\frac{\partial u_2}{\partial \bar{\theta}}+f_{2,u_3}\frac{\partial
u_3}{\partial \bar{\theta}}-\frac{\partial u_1}{\partial \bar{\theta}})\bigg{]}=0.\label{s5p}
\end{align}
Given some initial data $a_0$ and $b_0$, the definition (\ref{eq:def}) gives out $u_2$ and $u_3$. With these $a_0$, $b_0$, $u_2$ and $u_3$, we can solve Eq.~(\ref{s5}) for $u_1$. Then plug this information into Eqs.~(\ref{s1}) to (\ref{s4}) we can get $a_0$, $b_0$, $u_2$ and $u_3$ at next time step. Iterating this process, we can get the solutions $a_0$, $b_0$, $u_1$, $u_2$ and $u_3$. And it can be checked straightforwardly that the solutions $a_0$ and $b_0$ satisfy original equations (\ref{eq:F12}) and (\ref{eq:F345}). So with these coefficients $a_0$, $a_1$, $K$, $b_0$ and $b_1$, the coordinate transformations (\ref{ctu})-(\ref{ctphi}) gives us an new BS coordinate, with which we have following solutions of the coupled Einstein-Maxwell equations
\begin{eqnarray}
&&\gamma=\frac{c(u,\theta)}{r}+\frac{C(u,\theta)-\frac{1}{6}c(u,\theta)^3}{r^3}+\frac{\gamma_4(u,\theta)}{r^4}+\cdot\cdot\cdot\\
&&\beta=-\frac{c^2}{4r^2}-\frac{(A_2)^2+6c(C-\frac{1}{6}c^3)}{8r^4}\nonumber\\
&&\ \ \ \ \ \ \ \ \ \ \ +\frac{c(A_2)^2-(A_3)\cdot A_2-4c\gamma_4}{5r^5}+\cdot\cdot\cdot\\
&&U=L(u,\theta)-\frac{[\frac{\partial c}{\partial\theta}+2c\cot\theta]}{r^2}+\frac{U_3}{r^3}+\frac{U_4}{r^4}+\cdot\cdot\cdot\\
&&V=-\frac{1}{3}\Lambda
r^3+(L\cot\theta+\frac{\partial L}{\partial\theta})r^2+(1+\frac{1}{2}\Lambda c^2)r\nonumber\\
&&\ \ \ \ \ \ \ -2M(u,\theta)+\frac{V_1(u,\theta)}{r}+\cdot\cdot\cdot\\
&&F_{12}=\frac{A_2(u,\theta)}{r^2}+\frac{A_3(u,\theta)}{r^3}+\frac{A_4(u,\theta)}{r^4}+\cdot\cdot\cdot\\
&&F_{01}=\frac{B_2(u,\theta)}{r^2}+\frac{1}{r^3}\bigg{[}L\cdot
A_3+\frac{\partial
A_2}{\partial\theta}+A_2\cdot\cot\theta\bigg{]}\nonumber\\
&&\ \ \ \ \ \ \ \
+\frac{B_4(u,\theta)}{r^4}+\cdot\cdot\cdot\nonumber
\end{eqnarray}
In the following we will use the bared coordinate. In order to simplify the expression, we have dropped bar notation. In addition, we have changed the quantities $h$ and $V_0$ to
\begin{eqnarray}
h(u,\theta)=C(u,\theta)-\frac{1}{6}c(u,\theta)^3,\hspace{2mm}
V_0(u,\theta)=-2M(u,\theta),\nonumber
\end{eqnarray}
which makes our results easier to compare with the existing results in the literature.

For the coefficients in the above solutions, we have the following relation.
\begin{eqnarray}
&&U_3(u,\theta)=2N(u,\theta)+3c \frac{\partial c}{\partial \theta}+4c^2\cot\theta\nonumber\\
&&U_4(u,\theta)=\frac{3}{2}\frac{\partial
C}{\partial\theta}+3C\cdot\cot\theta-4c^2\frac{\partial
c}{\partial\theta}-4c^3\cot\theta\nonumber\\
&&\ \ \ \ \ \ \ \ \ \ \ \ \ \ -3cN-(A_2)\cdot B_2
+(A_2)^2L\nonumber\\
&&V_1(u,\theta)=-\big{[}\frac{\partial
N}{\partial\theta}+N\cot\theta-(\frac{\partial
c}{\partial\theta})^2-4c\frac{\partial
c}{\partial\theta}\cot\theta-\frac{1}{2}c^2\nonumber\\
&&\hspace{8mm}-4c^2\cot^2\theta -(B_2)^2-\frac{3}{8}\Lambda
c^4+\frac{3}{2}\Lambda c C\nonumber\\&&\hspace{8mm}+2A_2\cdot B_2 L
-L^2(A_2)^2+\frac{1}{4}\Lambda (A_2)^2\big{]}\nonumber\\
&&V_2(u,\theta)=-\frac{1}{2}\frac{\partial^2
C}{\partial\theta^2}-\frac{3}{2}\frac{\partial
C}{\partial\theta}\cot\theta +C-3N\frac{\partial
c}{\partial\theta}\nonumber\\
&&\hspace{4mm}-6cN\cot\theta-4c(\frac{\partial
c}{\partial\theta})^2-12c^2\frac{\partial
c}{\partial\theta}\cot\theta\nonumber\\
&&\hspace{4mm}\ -8c^3\cot^2\theta+\frac{\partial
A_2}{\partial\theta}B_2+A_2 B_2\cot\theta-LA_2\frac{\partial
A_2}{\partial\theta}\nonumber\\
&&\hspace{4mm} +\frac{1}{5}\Lambda
c(A_2)^2-L(A_2)^2\cot\theta-\frac{1}{5}\Lambda
(A_2)A_3\nonumber\\
 &&B_4(u,\theta)=\frac{1}{2}\frac{\partial
A_3}{\partial\theta}+\frac{1}{2}A_3\cot\theta-\frac{1}{2}c^2
B_2+LA_4\nonumber\\
&&\hspace{4mm}+\frac{1}{2}L c^2A_2  -3c
A_2\cot\theta-2A_2\frac{\partial c}{\partial\theta}-c\frac{\partial
A_2}{\partial\theta}
\end{eqnarray}
 and $N(u,\theta)$ is an arbitrary function.

Regarding to the components $F_{02}$ of the Maxwell field, we solve Eq.~$(\ref{maxwell3})$ and get
\begin{eqnarray}
F_{02}=C_0(u,\theta)+\frac{C_1(u,\theta)}{r}+\frac{C_2(u,\theta)}{r^2}+\cdot\cdot\cdot
\end{eqnarray}
where
\begin{eqnarray}
&&C_1(u,\theta)=c C_0-\frac{1}{2}\frac{\partial
B_2}{\partial\theta}+L\cdot
A_2\cot\theta+\frac{1}{2}A_2\frac{\partial
L}{\partial\theta}\nonumber\\
&& \ \ \ \ \ \ \ -\frac{1}{6}\Lambda A_3+\frac{1}{2}L\frac{\partial
A_2}{\partial\theta}+\frac{1}{3}\Lambda c A_2\\
&&C_2(u,\theta)=\frac{3}{4}A_2-\frac{1}{2}B_2\frac{\partial
c}{\partial\theta}-B_2\cdot
c\cdot\cot\theta-\frac{1}{4}\frac{\partial^2A_2}{\partial\theta^2}\nonumber\\
&&\  \ \ \ \  -\frac{1}{4}\frac{\partial
A_2}{\partial\theta}\cot\theta+\frac{1}{4}A_2\cot^2\theta-\frac{1}{4}c\frac{\partial
B_2}{\partial\theta}+\frac{1}{2}c^2\cdot C_0\nonumber\\
&&\ \  \ \ \  +L cA_2\cot\theta+\frac{3}{4}L\cdot
A_3\cot\theta+\frac{5}{12}\Lambda  A_2
c^2+\frac{1}{12}\Lambda c A_3\nonumber\\
&&\ \ \ \ \ \ \ -\frac{1}{4}c A_2\frac{\partial
L}{\partial\theta}-\frac{1}{6}\Lambda\cdot
A_4+\frac{1}{4}A_3\frac{\partial
L}{\partial\theta}+\frac{1}{4}cL\frac{\partial A_2}{\partial\theta}
\end{eqnarray}

Plugging above solutions $F_{01}$, $F_{02}$ and $F_{12}$ into Eq.~$(\ref{maxwell4})$ we obtain
\begin{eqnarray}
&&\frac{\partial A_2}{\partial u}=-\frac{1}{2}\frac{\partial B_2}{
\partial\theta}-c C_0-LA_2\cot\theta-\frac{1}{2}A_2\frac{\partial
L}{\partial\theta}\nonumber\\
&&\hspace{1cm}+\frac{1}{6}\Lambda A_3-\frac{1}{2}L\frac{\partial
A_2}{\partial\theta}-\frac{1}{3}\Lambda c A_2,\\
&&\frac{\partial A_3}{\partial u}=-\frac{1}{2}A_2+B_2\frac{\partial
c}{\partial\theta}+2B_2
c\cot\theta-\frac{1}{2}\frac{\partial^2A_2}{\partial\theta^2}\nonumber\\
&&\ \ \ -\frac{1}{2}\frac{\partial
A_2}{\partial\theta}\cot\theta+\frac{1}{2}A_2\cot^2\theta+\frac{1}{2}c\frac{\partial
B_2}{\partial\theta}-c^2 C_0\nonumber\\
&&\ \ \ -2LcA_2\cot\theta-\frac{3}{2}L A_3 \cot\theta
-\frac{5}{6}\Lambda c^2A_2-\frac{1}{6}\Lambda c
A_3\nonumber\\
&&\ \ \ +\frac{1}{2}cA_2\frac{\partial
L}{\partial\theta}+\frac{1}{3}\Lambda
A_4-\frac{3}{2}A_3\frac{\partial L}{\partial\theta}-\frac{1}{2}c
L\frac{\partial
A_2}{\partial\theta}\nonumber\\
&&\ \ \ -L\frac{\partial A_3}{\partial\theta}
\end{eqnarray}

Now we consider Eq.~(\ref{main4}), we get
\begin{eqnarray}
&&-\frac{2}{3}\Lambda c-L\cot\theta+\frac{\partial
L}{\partial\theta}=0,\\
&&\frac{\partial C}{\partial
u}=\frac{1}{12}\bigg{[}-24LC\cot\theta+3N\cot\theta-3\frac{\partial
N}{\partial\theta}-12L\frac{\partial
C}{\partial\theta}\nonumber\\
&&\ \ \ \ \ \ \ \ \  -12C\frac{\partial L}{\partial\theta}+4\Lambda
\gamma_4+6A_2 C_0+\Lambda(A_2)^2\bigg{]}. \label{C0}
\end{eqnarray}
from which we have
\begin{eqnarray}
c=\frac{3}{2\Lambda}(\frac{\partial
L}{\partial\theta}-L\cot\theta).\label{c}
\end{eqnarray}
By analyzing Eq.~$(\ref{maxwell1})$ we have
\begin{eqnarray}
&&\frac{\partial B_2}{\partial u}=C_0\cdot\cot\theta+\frac{\partial
C_0}{\partial\theta}-LB_2\cot\theta+\frac{1}{3}\Lambda
A_2\cot\theta\nonumber\\
&&\ \ \ \ -B_2\frac{\partial
L}{\partial\theta}+\frac{1}{3}\Lambda\frac{\partial
A_2}{\partial\theta}-\frac{1}{3}\Lambda Lc A_2+\frac{1}{6}\Lambda L
A_3+\frac{3}{2}LA_2\frac{\partial
L}{\partial\theta}\nonumber\\
&&\ \ \ \   -\frac{3}{2}L\frac{\partial B_2}{\partial\theta}
+A_2\frac{\partial L}{\partial u}+\frac{1}{2}L^2\frac{\partial
A_2}{\partial\theta}-c\cdot L\cdot C_0
\end{eqnarray}

Now it's time to consider the supplementary equations (\ref{supplambda}). Expanding them explicitly, we get
\begin{eqnarray}
&&R_{00}-8\pi T_{00}=\Lambda g_{00}\hspace{2mm}\Rightarrow\nonumber\\
&&\frac{2V}{r}\frac{\partial^2\beta}{\partial u\partial r}-\frac{V}{2r^2}\frac{\partial^2V}{\partial r^2}
-\frac{V^2}{r^2}\frac{\partial^2\beta}{\partial^2r}-\frac{V^2}{r^3}\frac{\partial \beta}{\partial r}-\frac{V}{r^2}\frac{\partial\beta}{\partial r}
\frac{\partial V}{\partial r}\nonumber\\
&&-\frac{1}{r^2}(\frac{\partial V}{\partial u}-2V\frac{\partial
\beta}{\partial u}) -\frac{V}{2r^2}\frac{\partial U}{\partial
\theta}+\frac{1}{2r}\frac{\partial U}{\partial \theta}\frac{\partial
V}{\partial r}
-\frac{2U}{r^2}\frac{\partial V}{\partial \theta}\nonumber\\
&&-\frac{1}{2r}\frac{\partial U}{\partial r}\frac{\partial V}{\partial \theta}+\frac{2UV}{r^2}\frac{\partial \beta}{\partial \theta}
+2\frac{\partial \gamma}{\partial u}\frac{\partial U}{\partial \theta}
+\frac{\partial^2 U}{\partial u\partial \theta }+2(\frac{\partial \gamma}{\partial u})^2\nonumber\\
&&+\frac{1}{r}(2UV\frac{\partial^2 \beta}{\partial
r\partial\theta}+U\frac{\partial \beta}{\partial
\theta}\frac{\partial V}{\partial r}
+V\frac{\partial \beta}{\partial r}\frac{\partial U}{\partial \theta}+2U\frac{\partial \beta}{\partial r}\frac{\partial V}{\partial \theta})\nonumber\\
&&-\frac{2U}{r}\frac{\partial \gamma}{\partial r}\frac{\partial
V}{\partial \theta}-2U\frac{\partial^2\beta }{\partial
u\partial\theta }
-2\frac{\partial \beta}{\partial u}\frac{\partial U}{\partial \theta}+2U\frac{\partial^2 \gamma}{\partial u\partial\theta}\nonumber\\
&&+U\frac{\partial^2 U}{\partial \theta^2 } +(\frac{\partial
U}{\partial \theta})^2+2(\frac{\partial\gamma }{\partial\theta
}-\frac{\partial \beta}{\partial \theta})U
\frac{\partial U}{\partial \theta}+\frac{U}{r}\frac{\partial^2 V}{\partial r\partial\theta}\nonumber\\
&&+[2(\frac{\partial \beta}{\partial \theta})^2-2\frac{\partial \beta}{\partial\theta }
\frac{\partial \gamma}{\partial \theta}+\frac{\partial^2\gamma }{\partial \theta^2 }]U^2
-\cot\theta\bigg{(}2U\frac{\partial \beta}{\partial u}-\frac{\partial U}{\partial u}\nonumber\\
&&-2U\frac{\partial \gamma}{\partial u}-U\frac{\partial U}{\partial
\theta}
-U^2\frac{\partial \gamma}{\partial \theta}+\frac{UV}{2r^2}-\frac{U}{2r}\frac{\partial V}{\partial r}-\frac{UV}{r}
\frac{\partial \beta}{\partial r}\bigg{)}\nonumber\\
&&+r^2e^{2(\gamma-\beta)}\bigg{[}-2(\frac{\partial^2\gamma }{\partial u\partial r }+\frac{1}{r}\frac{\partial \gamma}{\partial u})U^2
-2(\frac{\partial \gamma}{\partial u}-\frac{\partial \beta}{\partial u})U\frac{\partial U}{\partial r}\nonumber\\
&&-2U^2\frac{\partial^2 U}{\partial r\partial\theta
}-2U\frac{\partial U}{\partial r}\frac{\partial U}{\partial \theta}
-2U^3\frac{\partial^2 \gamma }{\partial r\partial\theta
}-\frac{2}{r}U^3\frac{\partial \gamma}{\partial\theta }\nonumber\\
&& -3U^2\frac{\partial \gamma}{\partial \theta}\frac{\partial
U}{\partial r}+2U^2\frac{\partial \beta}{\partial\theta
}\frac{\partial U}{\partial r}+\frac{UV}{r}\frac{\partial^2U
}{\partial r^2}
+\frac{4UV}{r^2}\frac{\partial U}{\partial r}\nonumber\\
&&+2(\frac{\partial \gamma}{\partial r}-\frac{\partial \beta}{\partial r})\frac{UV}{r}\frac{\partial U}{\partial r}
+\frac{U^2V}{r}\frac{\partial^2\gamma }{\partial r^2}+\frac{U^2}{r}\frac{\partial \gamma}{\partial r}\frac{\partial V}{\partial r }\nonumber\\
&&+\frac{U^2V}{r^2}\frac{\partial \gamma}{\partial r}-\frac{3U^2}{r}\frac{\partial U}{\partial \theta}-U^2\frac{\partial \gamma}{\partial r}
\frac{\partial U}{\partial\theta }+\frac{U^2}{r^2}\frac{\partial V}{\partial r}\nonumber\\
&&+\frac{V}{2r}(\frac{\partial U}{\partial r})^2-U^2(\frac{\partial
U}{\partial r}+\frac{U}{r}+U\frac{\partial \gamma}{\partial
r})\cot\theta-U\frac{\partial^2 U}{\partial u\partial r}\bigg{]}
\nonumber\\
&&+\frac{1}{2}r^4e^{4(\gamma-\beta)}U^2(\frac{\partial U}{\partial r})^2-\frac{1}{2r^3}e^{2(\beta-\gamma)}
\big{[}\frac{\partial^2V }{\partial\theta^2 }+2V\frac{\partial^2\beta }{\partial\theta^2 }\nonumber\\
&&+(2\frac{\partial \beta}{\partial \theta}-2\frac{\partial \gamma}{\partial\theta }+\cot\theta)
(\frac{\partial V}{\partial \theta}+2V\frac{\partial \beta}{\partial\theta })\big{]}+r^{-1}e^{-2\beta}VF_{01}^2\nonumber\\
&&-4e^{-2\beta}UF_{01}F_{02}+2r^{-2}e^{-2\gamma}F_{02}^2+r^2U^2e^{2\gamma-4\beta}F_{01}^2\nonumber\\
&&+2re^{-2\beta}UVF_{01}F_{12}-2r^2e^{2\gamma-4\beta}U^3F_{01}F_{12}\nonumber\\
&&-2r^{-3}e^{-2\gamma}VF_{12}F_{02}+2e^{-2\beta}U^2F_{12}F_{02}\nonumber\\
&&+r^{-4}e^{-2\gamma}V^2F_{12}^2-2r^{-1}e^{-2\beta}VU^2F_{12}^2+
r^2e^{2\gamma-4\beta}U^4F_{12}^2\nonumber\\
&&-\Lambda (Vr^{-1}e^{2\beta}-U^2r^2e^{2\gamma})=0,\label{supple1}
\end{eqnarray}
and
\begin{eqnarray}
&&R_{02}-8\pi T_{02}=0\hspace{2mm}\Rightarrow\nonumber\\
&& \frac{\partial^2\beta}{\partial
u\partial\theta}-\frac{\partial^2\gamma}{\partial
u\partial\theta}+2\frac{\partial \gamma }{\partial u}
\frac{\partial\gamma }{\partial\theta }-2\frac{\partial
\gamma}{\partial u
}\cot\theta-U\big{[}\frac{\partial^2\beta}{\partial\theta^2}\nonumber\\
&&+2(\frac{\partial\beta}{\partial\theta})^2
-2\frac{\partial\beta}{\partial\theta}\frac{\partial\gamma}{\partial\theta}+
\frac{\partial\beta}{\partial\theta}\cot\theta\big{]}-\frac{1}{2r}\frac{\partial^2
V}{\partial
r\partial\theta}\nonumber\\
&&+\frac{1}{2r^2}\frac{\partial V}{\partial \theta}
+(\frac{\partial\gamma}{\partial r}-\frac{\partial\beta}{\partial
r})\frac{1}{r}\frac{\partial V}{\partial\theta}
+r^2e^{2(\gamma-\beta)}\bigg{[}\frac{3}{2}U\frac{\partial^2U}{\partial r\partial\theta}\nonumber\\
&&+\frac{3U}{r}\frac{\partial
U}{\partial\theta}+2U(\frac{\partial^2\gamma}{\partial u\partial
r}+\frac{1}{r}\frac{\partial \gamma}{\partial u})+
\frac{1}{2}\frac{\partial ^2 U}{\partial u\partial
r}+2\frac{\partial^2\gamma}{\partial
r\partial\theta}U^2\nonumber\\
&&+(\frac{\partial\gamma}{\partial u} -\frac{\partial
\beta}{\partial u})\frac{\partial U}{\partial
r}+\frac{\partial\gamma}{\partial r}U\frac{\partial
U}{\partial\theta}+(2\frac{\partial\gamma}{\partial\theta}-\frac{\partial\beta}{\partial\theta})
U\frac{\partial U}{\partial r}\nonumber\\
&& +\frac{\partial U}{\partial r}\frac{\partial
U}{\partial\theta}-\frac{V}{2r}\frac{\partial ^2U}{\partial r^2}
-\frac{1}{r^2}(U\frac{\partial V}{\partial r}+2\frac{\partial U}{\partial r}V)\nonumber\\
&& -\frac{1}{r}\big{[}\frac{\partial^2\gamma}{\partial
r^2}UV+(\frac{\partial\gamma}{\partial
r}-\frac{\partial\beta}{\partial r}) \frac{\partial U}{\partial
r}V+\frac{\partial\gamma}{\partial r}U\frac{\partial V}{\partial
r}\big{]}\nonumber\\
&&-\frac{UV}{r^2}\frac{\partial \gamma}{\partial
r}+\frac{2U^2}{r}\frac{\partial\gamma}{\partial\theta}+U(\frac{1}{2}
\frac{\partial U}{\partial r}+\frac{U}{r}+\frac{\partial
\gamma}{\partial
r}U)\cot\theta\bigg{]}\nonumber\\
&&-\frac{1}{2}r^4e^{4(\gamma-\beta)}U(\frac{\partial U}{\partial
r})^2+2e^{-2\beta}F_{01}F_{02}-r^2Ue^{2\gamma-4\beta}F_{01}^2\nonumber\\
&&-2r^{-1}e^{-2\beta}VF_{01}F_{12}+2r^2U^2e^{2\gamma-4\beta}F_{01}F_{12}-\Lambda
U r^2e^{2\gamma}
\nonumber\\
&&+r^{-1}UVe^{-2\beta}F_{12}^2-r^2U^3e^{2\gamma-4\beta}F_{12}=0.\label{supple2}
\end{eqnarray}

The $r^0$ order term of $(\ref{supple2})$ leads to
\begin{eqnarray}
&&\- A_2L^3 =0,
\end{eqnarray}
and the $r^{-1}$ order term of $(\ref{supple2})$ leads to
\begin{eqnarray}
&&\- A_3L^3 =0.
\end{eqnarray}
The $r^{-1}$ order term of $(\ref{supple1})$ gives us
\begin{eqnarray}
\frac{9L}{8\Lambda}(\frac{\partial
L}{\partial\theta}-L\cot\theta)\frac{\partial}{\partial\theta}(\frac{\partial
L}{\partial\theta}-L\cot\theta)=0
\end{eqnarray}
Recalling the relation (\ref{c}), this equations can be reduced to
\begin{eqnarray}
\frac{9L}{8\Lambda}c\frac{\partial c}{\partial\theta}=0.
\end{eqnarray}
So we have $L=0$ or $\partial
c/\partial\theta=0$.

If $L=0$ we have $c=0$.

If $\partial
c/\partial\theta=0$, we have
\begin{eqnarray}
\frac{\partial^2L}{\partial \theta^2}-\frac{\partial L}{\partial
\theta}\cot\theta+\frac{L}{\sin^2\theta}=0.
\end{eqnarray}
The solution of this equation reads as
\begin{eqnarray}
L(u,\theta)=\xi(u)\sin\theta+\eta(u)\sin\theta\cdot\ln(\tan\frac{\theta}{2})
\end{eqnarray}
with two integration constants $\xi(u)$ and $\eta(u)$. The regularity of $L$ on a specific sphere of constant $u$ requires $\eta(u)=0$, so we have
\begin{eqnarray}
L(u,\theta)=\xi(u)\sin\theta.
\end{eqnarray}
Together with Eq.(\ref{c}) this also results in $c=0$.

So we can summarize
the solution as
\begin{eqnarray}
&&\gamma=\frac{C(u,\theta)}{r^3}+\frac{\gamma_4(u,\theta)}{r^4}+\cdot\cdot\cdot\\
&&\beta=-\frac{(A_2)^2}{8r^4}-\frac{ A_2A_3}{5r^5}+\cdot\cdot\cdot\\
&&U=L(u,\theta)+\frac{2N(u,\theta)}{r^3}\nonumber\\
&&\ \ \ \ \ \ \ \ \ \
+\frac{1}{r^4}\bigg{[}\frac{3}{2}\frac{\partial
C}{\partial\theta}-A_2 B_2+3C\cot\theta\bigg{]}+\cdot\cdot\cdot\\
&&V=-\frac{1}{3}\Lambda r^3+(L\cot\theta+L_2)r^2+r
-2M(u,\theta)\nonumber\\
&&\ \ \ \  -\frac{1}{r}\bigg{[}\frac{\partial N}{\partial\theta}+N\cot\theta-(B_2)^2+\frac{1}{4}\Lambda(A_2)^2\bigg{]}+\cdot\cdot\cdot\\
&&F_{12}=\frac{A_2(u,\theta)}{r^2}+\frac{A_3(u,\theta)}{r^3}+\frac{A_4(u,\theta)}{r^4}+\cdot\cdot\cdot\\
&&F_{01}=\frac{B_2(u,\theta)}{r^2}+\frac{1}{r^3}\bigg{[}\frac{\partial
A_2}{\partial\theta}+A_2\cot\theta\bigg{]}+\cdot\cdot\cdot\nonumber\\
&&F_{02}=C_0(u,\theta)-\frac{1}{r}\bigg{[}\frac{1}{6}\Lambda A_3
+\frac{1}{2}\frac{\partial
B_2}{\partial\theta}\bigg{]}+\frac{1}{r^2}\bigg{[}\frac{3}{4}A_2\nonumber\\
&&-\frac{1}{4}\frac{\partial^2A_2}{\partial\theta^2}-\frac{1}{4}\frac{\partial
A_2}{\partial\theta}\cot\theta+\frac{1}{4}A_2\cot^2\theta-\frac{1}{6}\Lambda
A_4 \bigg{]}+\cdot\cdot\cdot\nonumber
\end{eqnarray}
and
\begin{eqnarray}
&&\frac{\partial A_2}{\partial u}=\frac{1}{6}\Lambda
A_3-\frac{1}{2}\frac{\partial
B_2}{\partial\theta},\nonumber\\
&&\frac{\partial C}{\partial
u}=-\frac{1}{4}N\cot\theta+2LC\cot\theta-\frac{1}{12}\Lambda
(A_2)^2-\frac{1}{3}\Lambda \gamma_4\nonumber\\
&&\ \ \ \ \ \ \ \  \ \ +C\frac{\partial
L}{\partial\theta}-\frac{1}{2}A_2 C_0+\frac{1}{4}\frac{\partial
N}{\partial\theta}+L\frac{\partial C}{\partial\theta},\nonumber\\
&&\frac{\partial B2}{\partial u}=C_0\cot\theta+\frac{\partial
C_0}{\partial\theta}-LB_2\cot\theta+\frac{1}{3}\Lambda
A_2\cot\theta\nonumber\\
&&\ \ \ \ \ \ \ \ \ \ -B_2\frac{\partial
L}{\partial\theta}+\frac{1}{3}\Lambda\frac{\partial
A_2}{\partial\theta}-\frac{3}{2}L\frac{\partial
B_2}{\partial\theta},\nonumber\\
&&L=\xi(u)\sin\theta.
\end{eqnarray}
Submitting above solutions into Eqs.~(\ref{supple1}) and (\ref{supple2}), we can get
\begin{eqnarray}
&&\frac{\partial M}{\partial u}=-(C_0)^2-\frac{1}{2}(A_2)^2-\frac{3}{2}ML\cot\theta+\frac{1}{2}N\Lambda\cot\theta\nonumber\\
&&\ \ \ \ \ \ \ \ -\frac{1}{3}A_2\Lambda
C_0+\frac{1}{2}\Lambda\frac{\partial
N}{\partial\theta}-\frac{3}{2}M\frac{\partial
L}{\partial\theta}-L\frac{\partial M}{\partial\theta}\nonumber\\
&&\ \ \ \ \ \ \ +\frac{1}{3}L\Lambda B_2 A_4+L^4 A_4,
\end{eqnarray}
and
\begin{eqnarray}
&&\frac{\partial N}{\partial u}=\frac{1}{3}\bigg{[}-\frac{\partial
M}{\partial\theta}-4L\frac{\partial C}{\partial
u}-4L^2\frac{\partial C}{\partial\theta}-4N\frac{\partial
L}{\partial\theta}-L^3A_4
\nonumber\\
&&\ \ \ \ \ \ \ \ \ -4L\frac{\partial
N}{\partial\theta}+\frac{4}{3}L\Lambda \gamma_4+2C_0
B_2-4LC\frac{\partial L}{\partial\theta}+\Lambda\frac{\partial
C}{\partial\theta}\nonumber\\
&&\ \ \ \ \ \  \ \ \ -4LN\cot\theta+2\Lambda
C\cot\theta-8L^2C\cot\theta\bigg{]}.
\end{eqnarray}

\begin{figure}
\begin{tabular}{c}
\includegraphics[width=0.35\columnwidth]{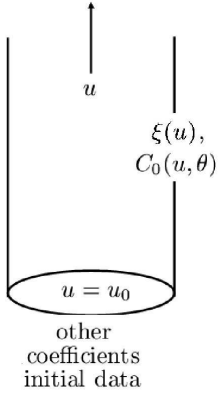}
\end{tabular}
\caption{Evolution structures for the coupled Einstein-Maxwell equations with cosmological constant. $u$
represents the Bondi time. Each cut of the cylinder represents the celestial
sphere at some time $u$. Here the Bondi's out going boundary
condition (\ref{eq:gamma2}) and (\ref{eq:f12}) for the gravitational field and electromagnetic field respectively.
$\xi(u)$ is the coefficient of variable $U$
and $C_0(u,\theta)$ is the coefficient of variable $F_{02}$.}\label{fig1}
\end{figure}

These results imply a characteristic structure which is summarized in Fig.~\ref{fig1}. This characteristic structure is different to the one found in \cite{BonVanMet62,Sac62}. The difference include two aspects. One is that the freedom $\xi$ for gravitational field is independent of the direction of the gravitational wave source locating in the celestial sphere. Instead it can only vary along the Bondi time $u$. The second aspect is that the freedom $\xi(u)$ is related to the BS metric variable $U$. While in \cite{BonVanMet62,Sac62} the freedom comes from the variable $\gamma$ (check  Sec.~\ref{sec::II}).

Using the quasi-Kinnersley tetrad (\ref{eq:QKtetradl})-(\ref{eq:QKtetradm}) we get the Newman-Penrose scalar $\Psi_4$
for gravitational wave
\begin{align}
\Psi_4=-\frac{C\Lambda^2}{6r}+\cdot\cdot\cdot,\label{Psi41}
\end{align}
and the Newman-Penrose scalar $\Phi_2=F_{ab}\bar{m}^an^b$ for electromagnetic wave
\begin{eqnarray}
\Phi_2=-\frac{C_0}{\sqrt{2}r}+\cdot\cdot\cdot\label{Phi21}
\end{eqnarray}
We can see that the electromagnetic wave behaves
the same as the case without cosmological constant. But the gravitational wave behaves totally different to the case without cosmological constant. Especially, when $\Lambda$ goes to zero, the gravitational wave here tends to zero also. The result can not recover the picture of the Einstein equations without cosmological constant. This is contradict to physical intuition.

\section{Boundary condition analysis} \label{sec::IV}

In the last section we have analyzed the coupled Einstein-Maxwell equations with Bondi's original out going boundary condition. The behavior of gravitational wave shows some bizarrerie. This reminds us the subtle results found by Ashtekar and his coworkers in \cite{AshBonKes15}. Ashtekar used conformal completion technique to investigate the Einstein equations with positive cosmological constant. The authors required the solutions of the Einstein equations asymptotically go to de Sitter space. They found that the boundary condition which requires the magnetic part of the gravitational field vanish at the null infinity leads to non-existence of gravitational waves. In the last section, we set our problem differently. We use Bondi's out going boundary condition to determine the solution of the coupled Einstein-Maxwell equations. We do not concern whether the solution is conformal completable or not. We do not concern whether the solution asymptotically goes to de Sitter space or not.

But the subtle results for gravitational wave got in the last section motivate us to analyze the boundary condition for the Einstein equations with non-vanishing cosmological constant in more detail.

Let's consider a model wave equation with the form as following
\begin{eqnarray}
\frac{\partial^2 Q}{\partial t^2}=\nabla^2Q-\Lambda Q.
\label{eq:wave1}
\end{eqnarray}
Here we use $\nabla^2$ to denote the usual Laplace operator on Euclidian space.
In order to analyze the out going boundary condition, we introduce
an out going coordinate $u=t-r$ to mimic the Bondi time. Decomposing $Q$ into spherical
harmonic components $Q_n$, we can get
\begin{eqnarray}
2(\frac{\partial^2 Q_n}{\partial r\partial
u}+\frac{1}{r}\frac{\partial Q_n}{\partial
u})&=&\frac{\partial^2Q_n}{\partial r^2}+\frac{2}{r}\frac{\partial
Q_n}{\partial r}-\frac{n(n+1)}{r^2}Q_n\nonumber\\
&&-\Lambda Q_n.
\end{eqnarray}
Expanding $Q_n=\sum_kL_n^k(u)r^{-k-1}$, we get the recurrence
relation
\begin{eqnarray}
2(k+1)\frac{dL_n^{k+1}}{du}=(n-k)(n+k+1)L^k_n-\Lambda
L_n^{k+2}.\label{eq:rec1}
\end{eqnarray}
Due to the requirement of convergence when $r$ goes to infinity,
$k\geq-1$. When $k=-1$, Eq.(\ref{eq:rec1}) gives us
\begin{eqnarray}
L_n^{-1}=\frac{\Lambda}{n(n+1)}L_n^1.\label{eq:Q0}
\end{eqnarray}
If $\Lambda=0$, we have $L_n^{-1}=0$. Consequently the out going
boundary condition for $\Lambda=0$ can be expressed as
\begin{eqnarray}
Q=\frac{Q^{(1)}}{r}+\cdot\cdot\cdot\label{eq:BDB}
\end{eqnarray}
This reminds us the Bondi's out going boundary condition (\ref{eq:gamma}).

If $\Lambda\neq0$, the out going boundary condition should
generally be expressed as
\begin{eqnarray}
Q=Q^{(0)}+\frac{Q^{(1)}}{r}+\cdot\cdot\cdot\label{eq:BDM}
\end{eqnarray}

As to the vacuum Einstein equation without cosmological constant,
the linearized equations about the flat metric $\eta_{\mu\nu}$ take
the wave equation form
\begin{align}
\Box\bar{h}_{\mu\nu}=0.
\end{align}
While for the
vacuum Einstein equation with cosmological constant, the
linearized equations take the form
\begin{align}
\Box\bar{h}_{\mu\nu}=\Lambda(\eta_{\mu\nu}+\frac{1}{2}h\eta_{\mu\nu})\label{eq:modellambda}
\end{align}
with $h=-\eta^{\mu\nu}\bar{h}_{\mu\nu}$. The corresponding full
metric is
$g_{\mu\nu}=\eta_{\mu\nu}+\bar{h}_{\mu\nu}+\frac{1}{2}h\eta_{\mu\nu}$. The equation form (\ref{eq:modellambda}) implies that the suitable out going boundary condition for the Einstein equation with cosmological constant should take the form (\ref{eq:BDM}) instead of the original Bondi's boundary condition.

Regarding to Maxwell equation, it always takes the form
\begin{align}
\Box A_{\mu}=0,
\end{align}
independent of $\Lambda$. So we conjecture that the out going boundary condition form (\ref{eq:f12}) is always valid for the Maxwell equations, no matter the Einstein equations the background metric related to include cosmological constant or not.

In the following section we will investigate the boundary condition (\ref{eq:BDM}) for the Einstein equations with cosmological constant.
\section{Apply the new boundary condition to the Einstein equations with $\Lambda$} \label{sec::V}
In this section we repeat the works done in Sec.~\ref{sec::III}. The difference is we apply here a new out going boundary condition instead of Bondi's original one (\ref{eq:gamma2}) to the Einstein equations with cosmological constant. Our new out going boundary condition is motivated by Eq.(\ref{eq:BDM}). We write it out specifically as
\begin{align}
\gamma=\Lambda
f(u,\theta)+\frac{c(u,\theta)}{r}+\frac{d(u,\theta)}{r^2}+\frac{h(u,\theta)}{r^3}+\cdot\cdot\cdot\label{gammanew}
\end{align}
for gravitational field and (\ref{eq:f12}) for Maxwell field.

Firstly we point out one important consequence resulted from above new boundary condition. This is about the geometry of the sphere with constant $u$ and constant $r$. In \cite{BonVanMet62,Sac62}, Bondi's original boundary condition makes the sphere go to standard sphere when $r$ goes to infinity. But now the new boundary condition (\ref{gammanew}) makes the sphere go asymptotically to
\begin{align}
e^{2\Lambda f}d\theta^2+e^{-2\Lambda f}\sin^2\theta d\phi^2.
\end{align}

Plugging above new boundary condition into Eq.~(\ref{maxwell2}) we get
\begin{eqnarray}
\frac{\partial A_1(u,\theta)}{\partial\theta}+A_1\cot\theta-2\Lambda
A_1\frac{\partial f}{\partial\theta}=0,
\end{eqnarray}
which corresponds to Eq.~(\ref{com_1}) in Sec.~\ref{sec::III}.
By integrating it, we get
\begin{eqnarray}
A_1(u,\theta)=\frac{\Sigma(u)e^{2\Lambda f}}{\sin\theta}
\end{eqnarray}
where $\Sigma(u)$ is the integration constant. But the
regularity condition of $A_1$ on sphere leads
$A_1(u,\theta)=0$.
Then the main equation $(\ref{main1})$ gives us
\begin{eqnarray}
&&\beta=H(u,\theta)-\frac{c^2}{4r^2}-\frac{2cd}{3r^3}\nonumber\\
&&\ \ \ \ \ \ -\frac{(A_2)^2e^{-2\Lambda
f}+6ch+4d^2}{8r^4}+\cdot\cdot\cdot\label{betanew}
\end{eqnarray}
And the equation $(\ref{main2})$ gives
\begin{eqnarray}
&&U=L(u,\theta)+\frac{2e^{2H-2\Lambda f}}{r}\frac{\partial
H}{\partial\theta}+\cdot\cdot\cdot,\label{unew}\\
&&\frac{\partial d(u,\theta)}{\partial\theta}+2
d(u,\theta)\cot\theta-2 \Lambda d \frac{\partial
f}{\partial\theta}=0.
\end{eqnarray}
The second of above equations leads
\begin{eqnarray}
d(u,\theta)=\frac{\Delta(u)e^{2\Lambda f}}{\sin^2\theta}
\end{eqnarray}
where $\Delta(u)$ is the integration constant. But again the
regularity condition of $d$ on sphere results in
$d(u,\theta)=0$.

The main Eq.~(\ref{main3}) gives
\begin{eqnarray}
&&V=-\frac{\Lambda}{3}e^{2H} r^3+(L\cot\theta+L_2)r^2+e^{2H-2\Lambda
f} \bigg{[}1\nonumber\\&&\ \ \ \ \ +4(\frac{\partial
H}{\partial\theta})^2+2\frac{\partial^2H}{\partial\theta^2}+2\frac{\partial
H}{\partial \theta}\cot\theta+\frac{1}{2}e^{2\Lambda f}\Lambda
c^2\nonumber\\
&&\ \ \ \ \ -4\Lambda\frac{\partial H}{\partial\theta}\frac{\partial
f}{\partial\theta}+3\Lambda\frac{\partial
f}{\partial\theta}\cot\theta+\Lambda\frac{\partial^2f}{\partial\theta^2}-2\Lambda^2(\frac{\partial
f}{\partial\theta})^2\bigg{]}r
\nonumber\\
&&\ \ \ \ \ \ +V_0(u,\theta)+\cdot\cdot\cdot\label{vnew}
\end{eqnarray}

Substituting Eqs.~(\ref{gammanew}), (\ref{betanew}), (\ref{unew}) and
(\ref{vnew}) to Eq.~(\ref{eq:BSmetric}), we get
\begin{eqnarray}
&&g_{00}=-[e^{2H}V^{(3)}-e^{2\Lambda
f}L^2]r^2-[e^{2H}V^{(2)}\nonumber\\
&&\hspace{10mm}-2U_1 e^{2\Lambda f} L-2e^{2\Lambda
f}L^2c]r+\cdot\cdot\cdot\nonumber\\
&&g_{01}=-e^{2H}+\frac{e^{2H}c^2}{2r^2}+\cdot\cdot\cdot\\
&&g_{02}=-e^{2\Lambda f}Lr^2-e^{2\Lambda f}[U_1+2Lc]r+\cdot\cdot\cdot\\
&&g_{22}=e^{2\Lambda f}r^2+2e^{2\Lambda f}cr+\cdot\cdot\cdot
\end{eqnarray}
where
\begin{eqnarray}
&&V^{(3)}=-\frac{\Lambda}{3}e^{2H}\nonumber\\
&&V^{(2)}=L\cot\theta+L_2\nonumber\\
&&U_1=2e^{2H-2\Lambda f}\frac{\partial H}{\partial\theta}\nonumber
\end{eqnarray}
In the following we will show that specific coordinate
transformation exists which preserves the BS metric form and at the same time makes $H$ and $L$ vanish. Starting from some coordinate system with non zero $H$ and $L$, we assume the following coordinate transformation
\begin{eqnarray}
&&u=a_0(\bar{u},\bar{\theta})+\frac{a_1(\bar{u},\bar{\theta})}{\bar{r}}+\cdot\cdot\cdot,\\
&&r=K(\bar{u},\bar{\theta})\bar{r}+\rho_0(\bar{u},\bar{\theta})+\cdot\cdot\cdot,\\
&&\theta=b_0(\bar{u},\bar{\theta})+\frac{b_1(\bar{u},\bar{\theta})}{\bar{r}}+\cdot\cdot\cdot,\\
&&\phi=\bar{\phi}.
\end{eqnarray}
To preserve the BS metric form (\ref{eq:BSmetric}), we
need
\begin{eqnarray}
&&\hspace{30mm}\bar{g}_{11}=0\hspace{2mm}\Rightarrow\hspace{2mm}\nonumber\\
&&[e^{2H}V^{(3)}-e^{2\Lambda f}L^2]K^2(a_1)^2-2e^{2H}K\cdot
a_1\nonumber\\
&&+2e^{2\Lambda f}LK^2a_1\cdot b_1-e^{2\Lambda f}K^2(b_1)^2=0,\label{cL1}\\
&&\hspace{30mm}\bar{g}_{12}=0\hspace{2mm}\Rightarrow\hspace{2mm}\nonumber\\
&&-[e^{2H}V^{(3)}-e^{2\Lambda f}L^2]K^2a_1\cdot
\frac{\partial a_0}{\partial\bar{\theta}}-e^{2\Lambda f}LK^2\cdot
a_1\cdot\frac{\partial b_0}{\partial\bar{\theta}}\nonumber\\
&&+e^{2H}K\cdot \frac{\partial a_0}{\partial\bar{\theta}}-e^{2\Lambda f}LK^2\cdot b_1\cdot
\frac{\partial a_0}{\partial\bar{\theta}}\nonumber\\
&&+e^{2\Lambda f}K^2b_1\cdot\frac{\partial b_0}{\partial\bar{\theta}}=0,\label{cL2}\\
&&\hspace{20mm}\bar{g}_{22}=e^{2\Lambda \bar{f}}\bar{r}^2+\mathcal{O}(\bar{r})\hspace{2mm}\Rightarrow\hspace{2mm}\nonumber\\
&&(e^{2H}V^{(3)}-e^{2\Lambda
f}L^2)K^2(\frac{\partial a_0}{\partial\bar{\theta}})^2+2e^{2\Lambda
f}LK^2\frac{\partial a_0}{\partial\bar{\theta}}\frac{\partial b_0}{\partial\bar{\theta}}\nonumber\\
&&-e^{2\Lambda f}K^2(\frac{\partial b_0}{\partial\bar{\theta}})^2=-e^{2\Lambda
\bar{f}},\label{cL3}\\
&&\hspace{15mm}\bar{g}_{33}=e^{-2\Lambda
\bar{f}}\bar{r}^2\sin^2\bar{\theta}+\mathcal{O}(\bar{r})\hspace{2mm}\Rightarrow\hspace{2mm}\nonumber\\
&&e^{-2\Lambda f}K^2\sin^2b_0=e^{-2\Lambda
\bar{f}}\sin^2\bar{\theta}\label{cL4}
\end{eqnarray}
Regarding to $H$ we have
\begin{eqnarray}
\bar{g}_{01}&=&(e^{2H}V^{(3)}-e^{2\Lambda
f}L^2)K^2\frac{\partial a_0}{\partial\bar{u}}a_1-e^{2H}\frac{\partial a_0}{\partial\bar{u}}K\nonumber\\
&&+Le^{2\Lambda
f}K^2\frac{\partial a_0}{\partial\bar{u}}b_1+Le^{2\Lambda f}K^2\frac{\partial b_0}{\partial\bar{u}}\cdot a1\nonumber\\
&&-e^{2\Lambda
f}K^2\frac{\partial b_0}{\partial\bar{u}}\cdot
b_1+\mathcal{O}(\frac{1}{\bar{r}^2})\nonumber\\
&=&-e^{2\bar{H}}+\mathcal{O}(\frac{1}{\bar{r}^2})
\end{eqnarray}
Here we have denoted the corresponding $H$ function in bared coordinate system with $\bar{H}$.
In order to make $\bar{H}=0$, we need
\begin{eqnarray}
1&=&e^{2\bar{H}}\nonumber\\
&=&-(e^{2H}V^{(3)}-e^{2\Lambda
f}L^2)K^2\frac{\partial a_0}{\partial\bar{u}}a_1+e^{2H}\frac{\partial a_0}{\partial\bar{u}}K\nonumber\\
&&-Le^{2\Lambda
f}K^2\frac{\partial a_0}{\partial\bar{u}}b_1-Le^{2\Lambda f}K^2\frac{\partial b_0}{\partial\bar{u}}\cdot a1\nonumber\\
&&+e^{2\Lambda
f}K^2\frac{\partial b_0}{\partial\bar{u}}\cdot
b_1.\label{cL5}
\end{eqnarray}
Regarding to $L$, we consider
\begin{eqnarray}
\bar{g}_{02}&=&-\big{[}(e^{2H}V^{(3)}-L^2e^{2\Lambda
f})K^2\frac{\partial a_0}{\partial\bar{u}}\frac{\partial a_0}{\partial\bar{\theta}}\nonumber\\
&&+Le^{2\Lambda f}K^2
\frac{\partial a_0}{\partial\bar{u}}\frac{\partial b_0}{\partial\bar{\theta}}+Le^{2\Lambda f}K^2\frac{\partial b_0}{\partial\bar{u}}\frac{\partial a_0}{\partial\bar{\theta}}\nonumber\\
&&-e^{2\Lambda f}K^2\frac{\partial b_0}{\partial\bar{u}}\frac{\partial b_0}{\partial\bar{\theta}}\big{]}\bar{r}^2+\mathcal{O}(\bar{r})\nonumber\\
&=&-e^{2\Lambda\bar{f}}\bar{L}\bar{r}^2+\mathcal{O}(\bar{r})
\end{eqnarray}
Here we have denoted the corresponding $f$ function and $L$ function in bared coordinate system with $\bar{f}$ and $\bar{L}$ respectively.
To make $\bar{L}=0$, we need
\begin{eqnarray}
&&(e^{2H}V^{(3)}-L^2e^{2\Lambda
f})K^2\frac{\partial a_0}{\partial\bar{u}}\frac{\partial a_0}{\partial\bar{\theta}}+Le^{2\Lambda f}K^2
\frac{\partial a_0}{\partial\bar{u}}\frac{\partial b_0}{\partial\bar{\theta}}\nonumber\\
&&+Le^{2\Lambda f}K^2\frac{\partial b_0}{\partial\bar{u}}\frac{\partial a_0}{\partial\bar{\theta}}-e^{2\Lambda f}K^2\frac{\partial b_0}{\partial\bar{u}}\frac{\partial b_0}{\partial\bar{\theta}}=0\label{cL6}
\end{eqnarray}

Firstly, we algebraically solve Eqs.~(\ref{cL3}) and (\ref{cL4}) for $K(\bar{u},\bar{\theta})$ and
$\bar{f}(\bar{u},\bar{\theta})$ in terms of $a_0$, $b_0$, $\frac{\partial a_0}{\partial\bar{\theta}}$ and $\frac{\partial b_0}{\partial\bar{\theta}}$. Then we substitute the solution $K(\bar{u},\bar{\theta})$ into Eqs.~(\ref{cL1}) and (\ref{cL2}), and algebraically solve them for $a_1$ and
$b_1$ in terms of $a_0$, $b_0$, $\frac{\partial a_0}{\partial\bar{\theta}}$ and $\frac{\partial b_0}{\partial\bar{\theta}}$. Looking Eqs.~(\ref{cL5}) and (\ref{cL6}) as an algebra equation system for $\frac{\partial a_0}{\partial\bar{u}}$ and $\frac{\partial a_0}{\partial\bar{u}}$, we solve them to get solutions with form
\begin{eqnarray}
\frac{\partial a_0}{\partial\bar{u}}=F_{a_0}(a_0,b_0,\frac{\partial a_0}{\partial\bar{\theta}},\frac{\partial b_0}{\partial\bar{\theta}},K,f,\bar{f},a_1,b_1),\\
\frac{\partial b_0}{\partial\bar{u}}=F_{b_0}(a_0,b_0,\frac{\partial a_0}{\partial\bar{\theta}},\frac{\partial b_0}{\partial\bar{\theta}},K,f,\bar{f},a_1,b_1).
\end{eqnarray}
After we substitute $K$, $\bar{f}$, $a_1$ and $b_1$ into above equations we can obtain a
Cauchy problem respect to two unknown functions $a_0$ and $b_0$,
from which we can get the evolution equations of $a_0$ and $b_0$
respectively. Choosing two smooth functions on sphere as initial data for $a_0$ and $b_0$, we can solve this Cauchy problem to get $a_0(\bar{u},\bar{\theta})$ and $b_0(\bar{u},\bar{\theta})$. Substitute them backwards, we can get $a_1(\bar{u},\bar{\theta})$, $b_1(\bar{u},\bar{\theta})$ and $K(\bar{u},\bar{\theta})$. With them we get the desired coordinate transformation. Here it is deserved to comment that this kind of coordinate transformation does not exist in general in Sec.~\ref{sec::III} to make $L=0$.

With this new coordinate system, we can summarize the solution for coupled Einstein-Maxwell equations as following
\begin{eqnarray}
&&\gamma=\Lambda f(u,\theta)+\frac{c(u,\theta)}{r}+\frac{C(u,\theta)-\frac{1}{6}c(u,\theta)^3}{r^3}\nonumber\\
&&\ \ \ \ \ \ \ +\frac{\gamma_4(u,\theta)}{r^4}+\frac{\gamma_5(u,\theta)}{r^5}+\cdot\cdot\cdot\\
&&\beta=-\frac{c^2}{4r^2}-\frac{(A_2)^2e^{-2\Lambda f}+6c(C-\frac{1}{6}c^3)}{8r^4}+\cdot\cdot\cdot\\
&&U=-\frac{e^{-2\Lambda f}(2c\cot\theta+\frac{\partial c}{\partial\theta}-2\Lambda c\frac{\partial f}{\partial\theta})}{r^2}\nonumber\\
&&\ \ \ \ \ \ \ \ \ \ \ \ +\frac{U_3(u,\theta)}{r^3}+\frac{U_4(u,\theta)}{r^4}+\cdot\cdot\cdot\\
&&V=-\frac{1}{3}\Lambda r^3+e^{2\Lambda f}[1+\frac{1}{2}e^{2\Lambda
f}\Lambda c^2+3\Lambda\frac{\partial f}{\partial\theta}
\cot\theta+\Lambda\frac{\partial^2f}{\partial\theta^2}\nonumber\\
&&\ \ \ \ \ \ \ -2\Lambda^2(\frac{\partial f}{\partial\theta})^2]r
-2M(u,\theta)+\frac{V_1(u,\theta)}{r}+\cdot\cdot\cdot\\
&&F_{12}=\frac{A_2(u,\theta)}{r^2}+\frac{A_3(u,\theta)}{r^3}+\cdot\cdot\cdot\\
&&F_{01}=\frac{B_2(u,\theta)}{r^2}+e^{-2\Lambda
f}\bigg{[}\frac{\partial
A_2}{\partial\theta}+(A_2)\cot\theta\nonumber\\
&&\ \ \ \ \ \ \ \ -2A_2\Lambda\frac{\partial
f}{\partial\theta}\bigg{]}\frac{1}{r^3} +\cdot\cdot\cdot\nonumber\\
&&F_{02}=C_0(u,\theta)+\frac{C_1(u,\theta)}{r}+\cdot\cdot\cdot
\end{eqnarray}
where
\begin{eqnarray}
&&U_3(u,\theta)=e^{-2\Lambda f}\big{[}2N(u,\theta)+3c\frac{\partial
c}{\partial\theta}+4c^2\cot\theta\big{]}\nonumber\\
&&U_4(u,\theta)=e^{-2\Lambda f}\bigg{[}-4c^3\cot\theta+3C
\cot\theta-4c^2\frac{\partial
c}{\partial\theta}\nonumber\\
&&\ \ \ \ \ \ \ \ +\frac{3}{2}\frac{\partial C}{\partial\theta}
-2\Lambda c^3\frac{\partial f}{\partial\theta}-3cN-3\Lambda C
\frac{\partial f}{\partial\theta}-A_2B_2 \bigg{]}\nonumber\\
&&C_1(u,\theta)=c C_0-\frac{1}{2}\frac{\partial
B_2}{\partial\theta}-\frac{1}{6}\Lambda
A_3+\frac{1}{3}\Lambda c A_2\nonumber
\end{eqnarray}
In above expressions, we have dropped the bar notation for simplicity. And we have set
\begin{eqnarray}
h(u,\theta)=C(u,\theta)-\frac{1}{6}c(u,\theta)^3,\hspace{2mm}
V_0(u,\theta)=-2M(u,\theta).\nonumber
\end{eqnarray}

Similar to Sec.~\ref{sec::III}, we plug the solutions for electromagnetic field into
Eq.~(\ref{maxwell4}). Then we obtain
\begin{eqnarray}
&&\frac{\partial A_2}{\partial u}=-\frac{1}{2}\frac{\partial B_2}{
\partial\theta}-c C_0
+\frac{1}{6}\Lambda A_3-\frac{1}{3}\Lambda c A_2,\\
&&\frac{\partial A_3}{\partial
u}=-\frac{1}{2}A_2+B_2\cdot\frac{\partial
c}{\partial\theta}+2B_2\cdot
c\cdot\cot\theta-\frac{1}{2}\frac{\partial^2A_2}{\partial\theta^2}\nonumber\\
&&\ \ \ -\frac{1}{2}\frac{\partial
A_2}{\partial\theta}\cot\theta+\frac{1}{2}A_2\cdot\cot^2\theta+\frac{1}{2}c\frac{\partial
B_2}{\partial\theta}-c^2\cdot C_0\nonumber\\
&&\ \ \ -\frac{5}{6}\Lambda c^2A_2-\frac{1}{6}\Lambda c
A_3+\frac{1}{3}\Lambda A_4
\end{eqnarray}
Regarding to $f$, the $r$ order term of the main Eq.~(\ref{main4}) gives us
\begin{eqnarray}
&&3\Lambda\frac{\partial f}{\partial u}-c\Lambda=0
\end{eqnarray}
which means
\begin{eqnarray}
\frac{\partial f}{\partial u}=\frac{1}{3}c(u,\theta).
\end{eqnarray}
and the $r^{-2}$ order term of the main Eq.~(\ref{main4}) gives us
\begin{eqnarray}
&&\frac{\partial C}{\partial u}=
\frac{1}{12}\bigg{[}6c^2\frac{\partial c}{\partial u}+4\Lambda
\gamma_4-12e^{-2\Lambda f}\Lambda c\frac{\partial
c}{\partial\theta}\frac{\partial
f}{\partial\theta}\nonumber\\
&&\ \ \ \ \ \ \ +6e^{-2\Lambda f}c^2\Lambda\cot\theta\frac{\partial
f}{\partial\theta}-e^{-2\Lambda f}6\Lambda c^2\frac{\partial^2
f}{\partial\theta^2}\nonumber\\
&&\ \ \ \ \ \ \ +e^{-2\Lambda f}\Lambda (A_2)^2-4c\Lambda C+6\Lambda
c^3\frac{\partial f}{\partial u}+12\Lambda
C\frac{\partial f}{\partial u}\nonumber\\
&&\  \ \ \ \ \ \ +6e^{-2\Lambda f}C_0 A_2-3e^{-2\Lambda
f}\frac{\partial N}{\partial\theta}+3e^{-2\Lambda
f}N\cot\theta\nonumber\\&&\ \ \ \ \ \ \ \ -2\Lambda
c^4+6cM\bigg{]}\label{C0}
\end{eqnarray}
The main Eq.~(\ref{maxwell1}) results in
\begin{eqnarray}
&&\frac{\partial B_2}{\partial u}=e^{-2\Lambda
f}\bigg{[}C_0\cdot\cot\theta+\frac{\partial
C_0}{\partial\theta}+\frac{1}{3}\Lambda A_2\cot\theta
+\frac{1}{3}\Lambda\frac{\partial A_2}{\partial\theta}\nonumber\\
&&\ \ \ \ \ \ \ \ \ \   \ \ \ \  \ \ \ \ \
-\frac{2}{3}A_2\cdot\Lambda^2\frac{\partial
f}{\partial\theta}-2C_0\cdot\Lambda\frac{\partial
f}{\partial\theta}\bigg{]}
\end{eqnarray}
The supplementary equations (\ref{supple1}) and (\ref{supple2}) lead to
\begin{eqnarray}
&&\frac{\partial M}{\partial u}=-\frac{1}{2}\bigg{[}2e^{-2\Lambda
f}\Lambda^2 N \frac{\partial
f}{\partial\theta}+\frac{4}{3}e^{-2\Lambda
f}c^2\Lambda^3(\frac{\partial f}{\partial\theta})^2
\nonumber\\
&&\hspace{3mm} +\frac{4}{3}e^{-2\Lambda f}\Lambda
c^2\cot^2\theta-\frac{2}{3}cC\Lambda^2-e^{-2\Lambda
f}N\Lambda\cot\theta\nonumber\\
&&\hspace{3mm}  +\frac{2}{3}e^{-2\Lambda f}\Lambda A_2\cdot C_0
-\frac{1}{2}e^{-4\Lambda f}\Lambda \frac{\partial
f}{\partial\theta}\cot\theta \nonumber\\
&&\hspace{3mm} -\frac{3}{2}e^{-4\Lambda f}\Lambda \frac{\partial
f}{\partial \theta}\cot^3\theta-14e^{-4\Lambda
f}\Lambda^3(\frac{\partial
f}{\partial\theta})^3\cot\theta\nonumber\\
&&\hspace{3mm} -6e^{-4\Lambda f}\Lambda^2(\frac{\partial
f}{\partial\theta})^2\cot^2\theta -4e^{-2\Lambda
f}\Lambda^2(\frac{\partial f}{\partial\theta})^2\frac{\partial
c}{\partial u}\nonumber\\
&&\hspace{3mm} -4e^{-2\Lambda f}\Lambda\frac{\partial^2c}{\partial
u\partial\theta}\frac{\partial f}{\partial\theta} +2e^{-2\Lambda
f}\Lambda \frac{\partial
c}{\partial u}\frac{\partial^2f}{\partial\theta^2}\nonumber\\
&&\hspace{3mm}  +2e^{-2\Lambda f}\Lambda\frac{\partial
c}{\partial\theta}\frac{\partial^2 f}{\partial
u\partial\theta}+2e^{-2\Lambda f}c\Lambda
\frac{\partial^3f}{\partial
u\partial\theta^2}\nonumber\\
&&\hspace{3mm}    +\frac{3}{2}e^{-4\Lambda f}\Lambda
\frac{\partial^2f}{\partial\theta^2}\cot^2\theta+5e^{-4\Lambda
f}\Lambda^2\frac{\partial
f}{\partial\theta}\frac{\partial^3f}{\partial\theta^3}\nonumber\\
&&\hspace{3mm}  -18e^{-4\Lambda
f}\Lambda^3\frac{\partial^2f}{\partial\theta^2}(\frac{\partial
f}{\partial\theta})^2+2e^{-2\Lambda f}c^2\Lambda-e^{-2\Lambda
f}\Lambda\frac{\partial
N}{\partial\theta}\nonumber\\
&&\hspace{3mm} -2e^{-4\Lambda f}\Lambda\frac{\partial^3
f}{\partial\theta^3}-\frac{4}{3}e^{-2\Lambda f}\Lambda
(\frac{\partial
c}{\partial\theta})^2+\frac{4}{9}c^4\Lambda^2\nonumber\\
&&\hspace{3mm}    +\frac{2}{3}e^{-2\Lambda
f}c\Lambda^2\frac{\partial c}{\partial\theta}\frac{\partial
f}{\partial\theta}-3e^{-2\Lambda f}c\Lambda \frac{\partial
c}{\partial\theta}\cot\theta\nonumber\\
&&\hspace{3mm} +\frac{2}{3}e^{-2\Lambda f}\Lambda A_2\cdot
B_2\frac{\partial c}{\partial\theta}+6e^{-2\Lambda
f}\Lambda\frac{\partial c}{\partial u}\frac{\partial
f}{\partial\theta}\cot\theta\nonumber\\
&&\hspace{3mm} +2e^{-2\Lambda f}\frac{\partial c}{\partial u}
+e^{2\Lambda f}(A_2)^2+2e^{-2\Lambda
f}c\Lambda\frac{\partial^2f}{\partial u\partial\theta}\cot\theta \nonumber\\
&&\hspace{3mm}+15e^{-4\Lambda f}\Lambda^2\frac{\partial
f}{\partial\theta}\frac{\partial^2
f}{\partial\theta^2}+8e^{-4\Lambda f}\Lambda^4(\frac{\partial
f}{\partial\theta})^4 \nonumber\\
&&\hspace{3mm} -3e^{-2\Lambda f}\frac{\partial^2c}{\partial
u\partial\theta}\cot\theta-\frac{1}{2}e^{-4\Lambda f}\Lambda
\frac{\partial^4f}{\partial\theta^4}+4e^{-4\Lambda
f}\Lambda\frac{\partial^2f}{\partial\theta^2}\nonumber\\
&&\hspace{3mm} +3e^{-4\Lambda
f}\Lambda^2(\frac{\partial^2f}{\partial\theta^2})^2-e^{-2\Lambda
f}\frac{\partial^3 c}{\partial
u\partial\theta^2}+2(\frac{\partial c}{\partial u})^2\nonumber\\
&&\hspace{3mm}-13e^{-4\Lambda f}\Lambda^2(\frac{\partial
f}{\partial\theta})^2+2e^{-2\Lambda
f}(C_0)^2-\frac{5}{3}e^{-2\Lambda f}c\Lambda
\frac{\partial^2c}{\partial\theta^2}\nonumber\\
&&\hspace{3mm} -4e^{-2\Lambda f}c\Lambda^2\frac{\partial
f}{\partial\theta}\frac{\partial^2f}{\partial
u\partial\theta}+\frac{4}{3}e^{-2\Lambda f}c\Lambda A_2
B_2\cot\theta\nonumber\\
&&\hspace{3mm}-\frac{4}{3}e^{-2\Lambda f}c\Lambda^2A_2\cdot
B_2\frac{\partial f}{\partial\theta}+\frac{4}{3}e^{-2\Lambda
f}c^2\Lambda^2\frac{\partial
f}{\partial\theta}\cot\theta\bigg{]}\nonumber\\
\end{eqnarray}
and
\begin{eqnarray}
&&\frac{\partial N}{\partial u}=\frac{1}{3}\bigg{[}4\Lambda
N\frac{\partial f}{\partial u}+5e^{-2\Lambda f}c\Lambda
\frac{\partial f}{\partial\theta }-4e^{-2\Lambda
f}c\Lambda^3(\frac{\partial f}{\partial
\theta})^3\nonumber\\
&&\ \ \ \ \  \ -e^{-2\Lambda f}c\Lambda \frac{\partial^3f
}{\partial\theta^3 }-\frac{\partial c}{\partial\theta}\frac{\partial
c}{\partial u}-4c\frac{\partial c}{\partial
u}\cot\theta+\Lambda\frac{\partial
C}{\partial\theta}\nonumber\\
&&\ \ \ \ \ \ \ \   -3c\frac{\partial^2c}{\partial
u\partial\theta}-\frac{4}{3}\Lambda c^3\cot\theta+2\Lambda
C\cot\theta-\frac{8}{3}\Lambda c^2\frac{\partial
c}{\partial\theta}\nonumber\\
&&\ \ \ \ \ \ \ \  -\frac{4}{3}\Lambda^2c^3\frac{\partial
f}{\partial\theta}-\frac{4}{3}\Lambda c N-2\Lambda^2C\frac{\partial
f}{\partial\theta}-4c^2\Lambda \frac{\partial^2f}{\partial
u\partial\theta}\nonumber\\
&&\ \ \ \ \ \ \ \  +2\Lambda c\frac{\partial f}{\partial
u}\frac{\partial c}{\partial\theta}-8c\Lambda \frac{\partial
c}{\partial u}\frac{\partial
f}{\partial\theta}+8c^2\Lambda^2\frac{\partial f}{\partial
u}\frac{\partial f}{\partial\theta}-\frac{\partial M}{\partial\theta}\nonumber\\
&&\ \ \ \ \ \ \ \  +6e^{-2\Lambda f}c\Lambda^2(\frac{\partial
f}{\partial\theta})^2\cot\theta+3e^{-2\Lambda f}c\Lambda
\frac{\partial f}{\partial\theta}\cot^2\theta\nonumber\\
&&\ \ \ \ \ \ \ +6e^{-2\Lambda f}c\Lambda^2\frac{\partial
f}{\partial\theta}\frac{\partial^2
f}{\partial\theta^2}-3e^{-2\Lambda
f}c\Lambda\frac{\partial^2f}{\partial^2\theta}\cot\theta +2B_2
C_0\bigg{]}\nonumber
\end{eqnarray}
respectively. When $\Lambda=0$, above two equations become
\begin{eqnarray}
&&\frac{\partial M}{\partial u}=-(\frac{\partial c}{\partial
u})^2+\frac{1}{2}\frac{\partial }{\partial
u}[\frac{\partial^2c}{\partial\theta^2}+3\frac{\partial
c}{\partial\theta}\cot\theta-2c]\nonumber\\
&&\ \ \ \ \ \ \ \ \ -2(C_0)^2-(A_2)^2,\\
&&\frac{\partial N}{\partial u}=\frac{1}{3}\bigg{[} -\frac{\partial
M}{\partial\theta}-\frac{\partial c}{\partial\theta}\frac{\partial
c}{\partial u}-4c\frac{\partial c}{\partial
u}\cot\theta\nonumber\\
&&\ \ \ \ \ -3c\frac{\partial ^2c}{\partial
u\partial\theta}+2B_2C_0\bigg{]}
\end{eqnarray}
They are consistent with the results in \cite{vander69,BieCheYau11}.

\begin{figure}
\begin{tabular}{c}
\includegraphics[width=0.5\columnwidth]{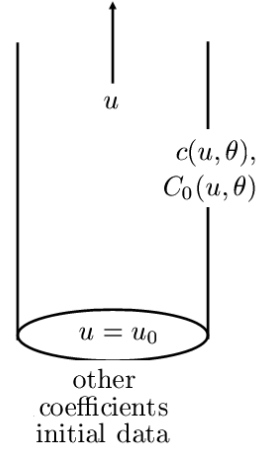}
\end{tabular}
\caption{The same to Fig.~\ref{fig1} but now the new out going boundary
condition (\ref{gammanew}) is used.
$c(u,\theta)$ is the coefficient of variable $\gamma$
and $C_0(u,\theta)$ is the coefficient of variable $F_{02}$. This evolution picture is the same to the one found in \cite{BonVanMet62,Sac62}.}\label{fig2}
\end{figure}

These results represent a characteristic structure which is different to Fig.~\ref{fig1}. We summarize this structure in Fig.~\ref{fig2}. Interestingly we can find this characteristic structure is roughly the same to the one found in \cite{BonVanMet62,Sac62}.

Based on the quasi-Kinnersley tetrad (\ref{eq:QKtetradl})-(\ref{eq:QKtetradm}) we can calculate the Newman-Penrose scalar $\Psi_4$ to get
\begin{eqnarray}
\Psi_4&=&\frac{1}{r}\left[-\frac{\partial^2c}{\partial
u^2}+\Lambda e^{-2 \Lambda f}(\frac{4}{3}c+\frac{2}{3}\cot^2\theta c-\frac{1}{6}\frac{\partial c}{\partial\theta}-\frac{1}{6}\frac{\partial^2 c}{\partial\theta^2})\right.\nonumber\\
&&+\Lambda^2e^{-2\Lambda f}(\frac{4c\cot\theta}{3}\frac{\partial
f}{\partial\theta}+\frac{4c}{3}\frac{\partial^2f}{\partial\theta^2}
-\frac{1}{3}\frac{\partial c}{\partial\theta}\frac{\partial f}{\partial\theta})\nonumber\\
&&\left.+\Lambda^2(\frac{2c^3}{3}-\frac{C}{6})-\Lambda^3\frac{2c}{3}e^{-2\Lambda
f}(\frac{\partial f}{\partial\theta})^2\right]+\cdot\cdot\cdot
\end{eqnarray}
We can see that when the novel outgoing boundary condition
(\ref{gammanew}) is considered, the gravitational wave's behavior is
similar to the result in Bondi's work. The major part
$-\frac{\partial^2c}{\partial u^2}$ is exactly the term given by
Bondi, when $\Lambda$ goes to zero, Bondi's original result is
recovered. Regarding to the Maxwell field the corresponding Newman-Penrose scalar
$\Phi_2$ has the form
\begin{eqnarray}
\Phi_2=-\frac{C_0}{\sqrt{2}r}.
\end{eqnarray}
This result is same as Eq.~(\ref{Phi21}) and is also the same to the one in
the case without cosmological constant. It means that the two
different boundary conditions for gravitational field discussed
previously do not affect the behavior of electromagnetic wave.

\section{Summary and discussion} \label{sec::VI}

Motivated by current cosmological observations, we have investigated the Einstein equations with cosmological constant with the method presented in Bondi's seminal work \cite{BonVanMet62,Sac62}. Since the cosmological constant is so tiny compared to realistic gravitational wave source, we are not sure how much effect of the cosmological constant may act on practical gravitational wave detection. But theoretical interests inspire us to
investigate the coupled Einstein-Maxwell equations which admit
a cosmological constant.

Mathematically, Bondi's original outgoing boundary
condition is among the assumptions in their seminal work \cite{BonVanMet62,Sac62}. Followed closely
Bondi's method, we firstly also took this assumption to analyze the behavior
of gravitational wave and electromagnetic wave and the
characteristic structure of coupled Einstein-Maxwell equations. We find that Bondi's ``news function", $c(u,\theta)$,
does not exist any more when cosmological constant is nonzero. The
amplitude of the gravitational wave is proportional to the square of
the cosmological constant. The characteristic structure of the dynamical fields is totally different to the one found in \cite{BonVanMet62,Sac62}.
Very recently, Ashtekar and his coworkers used conformal completion method to investigate this problem \cite{AshBonKes15}. They got very similar results to the ones got in current work. They generalized the boundary condition in $\Lambda=0$ case to $\Lambda>0$ case and found that ``the gravitational waves do not carry away (de Sitter) energy or momentum across null infinity!" This is consistent to our result (\ref{Psi41}). Our result implies that the gravitational wave is proportional to $\Lambda^2$. When $\Lambda$ is small, the gravitational wave is negligible.

The results rooted from Bondi's original outgoing boundary
condition seem quite unphysical. Then detail analysis about the boundary condition shows that Bondi's original outgoing boundary
condition is not consistent to the Einstein equations with a cosmological constant. The analysis also motivates us to propose a new out going boundary condition.

When the new out going boundary condition
is used, the gravitational wave's behavior is the same to the
results given by \cite{BonVanMet62,vander69}. Different to
gravitational wave, the electromagnetic wave's behavior does not
depend on these different type outgoing boundary conditions. But anyhow, both gravitational wave and electromagnetic wave behave well with our new out going boundary condition. And more when $\Lambda$ goes to zero, original result in \cite{BonVanMet62,Sac62} will recover naturally.

Although our new out going boundary condition is quite attractive, several interesting and important issues are left to study more in the near future. These issues include what kind of symmetry is reduced from this new boundary condition. This symmetry corresponds to the coordinate freedom related to the BS metric form in question. And related problem to this freedom is how to understand the gauge invariant property of the gravitational wave and the electromagnetic wave got in current work. Is it possible reexpress our new boundary condition in the conformal completion language and port it into the framework constructed in \cite{AshBonKes15}? All these interesting problems are out of the scope of current paper. We leave them to the on going work.

\acknowledgments
We thank Xiao Zhang and Xiaoning Wu for many useful discussions and
comments on the manuscript. This work was supported by the NSFC
(No.~11375260, and No.~11401199 ) and the Research
Foundation of Education Bureau of Hunan Province, China
(No.~14C0254).

\bibliography{refs}

\end{document}